\documentclass{aa}

\usepackage{graphics,graphicx}
\usepackage{xcolor}
\usepackage{amsmath}
\usepackage{amssymb}
\usepackage{braket}
\usepackage{wrapfig}
\usepackage{caption}
\usepackage{subcaption}
\usepackage{natbib}

\bibpunct{(}{)}{;}{a}{}{,} 

\begin{document}
\title{A quantum theory of the alignment and polarization of very small dust grains}
\titlerunning{Polarization of very small dust grains}
\author{Boy Lankhaar\inst{1} 
}
\institute{\inst{1}Institute of Theoretical Astrophysics, The Faculty of Mathematics and Natural Sciences, \\
University of Oslo, Sem Sælands vei 13, Oslo, Norway
\email{boy.lankhaar@astro.uio.no}}
\date{}

\abstract
{Anomalous microwave emission (AME) is a component of interstellar medium emission peaking at $\sim$10--60\,GHz. Its polarization is both a CMB foreground and a probe of the alignment physics of very small dust grains.}
{We quantify when the purely rotational electric-dipole emission from very small interstellar grains (spinning dust/AME) can become measurably polarized, and we quantify related UV/optical/IR polarization diagnostics.}
{We develop a quantum-mechanical symmetric-top model for an axisymmetric very small grain and express polarized emission and absorption coefficients in terms of irreducible density-matrix moments. Alignment is driven by anisotropic illumination; we solve a simplified two-manifold pumping model and compute (polarized) emission and absorption signatures for different dipole configurations and grain-size distributions.}
{Anisotropic illumination can generate polarized spinning-dust emission; in optimal geometries, polarization fractions near the emission peak reach the percent level, while more modest anisotropies reduce the polarization fraction. The same physics predicts polarized UV/optical/IR absorption that can be appreciable in strongly illuminated environments, whereas IR vibrational emission is predicted to be negligibly polarized. 
}
{Spinning-dust polarization depends on radiation anisotropy, dipole geometry, and the competition between alignment pumping and depolarization by rotational emission. Joint constraints from AME polarization and UV/optical/IR absorption polarimetry provide a direct test of alignment by anisotropic radiation fields and help bound polarized microwave foregrounds.}


\maketitle

\section{Introduction}
Anomalous microwave emission (AME) is a component of interstellar medium (ISM)
emission at centimetre wavelengths, peaking in the $\sim$10--60\,GHz range \citep{dickinson:18}. 
The leading physical interpretation of AME is that it originates from electric-dipole radiation from rapidly rotating very small grains (VSGs) \citep{draine:98,dickinson:18}. AME overlaps with the frequency bands used for cosmic microwave background (CMB) observations \citep{kogut:96, planck:11}. Also, because the relevant grains have effective sizes of only a few \AA, AME provides a potentially sensitive probe of the microphysics of the smallest end of the interstellar dust population.

For these reasons, AME modeling based on so-called spinning-dust theory has received sustained attention. Predicting the total emissivity requires modeling both the VSG's rotational state distribution and the radiation spectrum of a non-spherical rotor \citep{draine:98, ali:09, silsbee:11, zhang:25}. 
While most applications have focused on total intensity, AME polarization is also of considerable interest. It constitutes a potentially important foreground for CMB experiments and, more fundamentally, it may encode information about magnetic field structure and the physical conditions that control alignment in the smallest interstellar dust grains \citep{dickinson:18, herman:23, gonzalez:25, draine:03}.

Expectations for AME polarization have evolved alongside our understanding of alignment and dissipation in VSGs. Resonance paramagnetic relaxation was proposed as an efficient alignment mechanism for VSGs rotating at GHz frequencies, potentially enabling a polarized spinning-dust signal \citep{lazarian:00, hoang:13, hoang:16b}. However, quantum-mechanical considerations indicate that dissipative channels may be strongly suppressed in VSGs, effectively blocking alignment of the angular momentum with the magnetic field for the smallest grains responsible for AME \citep{draine:16}. In this limit, spinning-dust emission is expected to be negligibly polarized at microwave frequencies \citep{draine:16}. These considerations motivate exploring alignment pathways that do not rely on efficient internal dissipation in VSGs.


Anisotropic illumination offers such a pathway. Directional radiation fields can induce alignment, through mechanisms closely related to polarized line formation and molecular alignment in the interstellar medium \citep{landi:06, yan:06, lankhaar:20a}. For PAHs and likewise planar VSGs, similar effects have been explored in conjunction with resonance paramagnetic relaxation \citep{hoang:18}, or the context of polarized infrared emission features \citep{sironi:09}. Developing a self-consistent framework that extends these ideas to polarized spinning-dust emission and to complementary UV/optical/IR diagnostics is therefore warranted.

In this work we develop a quantum-mechanical formulation for the polarization of radiation from VSGs, with a focus on the conditions under which their purely rotational emission (spinning dust/AME) can become measurably polarized. We treat the VSG as an oblate symmetric top and describe alignment in terms of irreducible density-matrix moments, which provide a convenient bridge between the quantum description and the classical alignment factors commonly used in spinning-dust modeling. The central alignment mechanism that we investigate is anisotropic absorption of those photons that lead to intermittent heating events. The framework we derive delivers (coupled) predictions for polarization features of spinning-dust emission, UV/optical/IR absorption, and vibrational emission.


This paper is organized as follows. Section~2 introduces the physical model for
an axisymmetric VSG and summarizes the relevant rotational structure and
dipole properties. In Sect.~3 we formulate radiative transition rates in the
electric-dipole approximation. Section~4 develops the alignment formalism and
derives the steady-state alignment under anisotropic illumination. In Sect.~5 we
apply these results to the (polarized) emissivity of spinning VSGs. Section~6
discusses UV and IR polarization signatures.
We discuss implications and connections to
existing constraints in Sect.~7 and summarize our conclusions in Sect.~8.

\section{Physical model of a very small dust grain}
\subsection{Rotational structure of an axisymmetric nanograin}
We set out to model the energy structure of an interstellar VSG. In full generality, the VSG Hamiltonian, which is discussed in detail in Appendix A.1, contains rotational, vibrational, electronic, and nuclear spin degrees of freedom. Strictly, these degrees of freedom cannot be seen in isolation from each other, since they interact through Coriolis interactions (vibration-rotation), fine structure interactions (electron-rotation), hyperfine structure interactions (nuclear spin-rotation), as well as through vibronic coupling. 

Our focus here is on VSGs with effective sizes $\lesssim 15$ \AA. For particles of this size, under typical interstellar conditions, the rotational angular momentum is very large (see Eq.~\ref{eq:J_typ}), and greatly exceeds the total electronic spin, nuclear spin and vibrational angular momenta. When this is the case, the coupling between rotation and the internal vibrational, electronic, and nuclear spin structure is subdominant, compared to the rotational energy. Therefore, in good approximation, the rotational motion effectively decouples from the remaining degrees of freedom, allowing the total wavefunction to be written in product form,
$\Psi \approx \psi_{\mathrm{rot}} \psi_{\mathrm{rest}}$.

Having established this, we proceed to consider the rotational structure of the VSG. We approximate the VSG as an oblate symmetric top, for which the rotational energy levels are
\begin{subequations}
\begin{align}
E_{\mathrm{rot}} = hB J(J+1) - h(B-A)K^2,
\end{align}
where $h$ indicates Planck's constant, and where we approximate the rotational constants as those of a homogeneous oblate spheroid,
\begin{align}
A &= \frac{h}{8\pi I_{\parallel}} = \frac{15 h}{64 \pi^3} \frac{1}{\rho a^5 s}, \\
B &= \frac{h}{8\pi I_{\perp}} = \frac{15 h}{32 \pi^3} \frac{1}{\rho a^5 s (1+s^2)}.
\end{align}
\end{subequations}
Here, $I_{\parallel}$ and $I_{\perp}$ denote the moments of inertia parallel and perpendicular to the symmetry axis, $\rho$ is the VSG density, $a$ characterizes the VSG size along its long axis, and $s<1$ is the shape parameter describing the degree of oblateness.

The eigenstates of a symmetric top are described by the rotational angular momentum quantum number $J$ and its projection onto the body-fixed (grain) symmetry axis, $K$. This quantum number is analogous to the classical parameter $K/J\simeq\cos \theta_{\hat{\boldsymbol{a}}_1 \boldsymbol{J}} = \hat{\boldsymbol{J}} \cdot \hat{\boldsymbol{a}}_1$, with $\boldsymbol{J}$ the angular momentum vector and $\boldsymbol{a}_1$ the direction of the grain symmetry axis, that is employed in \citet{draine:16}. A third quantum number, $M$, describes the projection of angular momentum onto a space-fixed axis. We will later motivate that the proper choice of the space-fixed symmetry axis is the magnetic field direction. Then, this quantum number is analogous to the classical parameter, $M/J\simeq \cos \theta_{\boldsymbol{B} \boldsymbol{J}} = \hat{\boldsymbol{J}} \cdot \hat{\boldsymbol{B}}$, where $\boldsymbol{B}$ is the magnetic field, that is employed in \citet{draine:16}. Since the energy does not depend on $M$ in the absence of external fields, the levels exhibit a $(2J+1)$-fold degeneracy. Under the approximation that the rotational Hamiltonian is decoupled from the internal degrees of freedom, we need not explicitly detail the other interactions. Instead, we group all remaining non-rotational quantum numbers under the symbol $\gamma$. The total wavefunction is then expressed as the product state $\Psi = \ket{JKM} \ket{\gamma}$, which we denote as $\ket{(\gamma) JKM}$, treating $\gamma$ as a spectator quantum number.

\subsection{Thermal Cycling and Rotational State Redistribution}
Due to their small size, interstellar VSGs undergo stochastic heating. The absorption of a single quantum from the interstellar radiation field (ISRF) can raise the grain temperature to $\gtrsim 500\,\mathrm{K}$, followed by rapid cooling, on the order of seconds, to temperatures of the order of $10$s of Kelvin \citep{draine:01}. The cooling time can be contrasted to the ISRF absorption rate of a VSG, which is approximately \citep{draine:16}
\begin{align}
\label{eq:abs_rate}
\dot{N}_{\mathrm{abs}} \approx 5 \times 10^{-7} \ \mathrm{s}^{-1}\ 
U \left(\frac{a}{10 \ \mathrm{\text{\AA}}} \right)^{3},
\end{align}
where $U$ is the radiation field intensity relative to the local ISRF. Thus, grain heating occurs in brief ``thermal spikes'', while the grain spends the vast majority of its time in a cold vibrational state.

While the energy structure of rotational and vibrational motions in VSGs can be viewed in isolation, they are not necessarily uncoupled. Coriolis coupling allows exchange between rotational and vibrational angular momentum. This exchange conserves the total angular momentum of the grain, but redistributes its body-fixed projection, indicated by $K$. Whether such coupling channels are open depends on the density of vibrational states. As shown by \citet{draine:16}, above a characteristic internal temperature $T_{\rm int}$, rotational angular momentum efficiently couples to internal vibrational modes, rapidly driving the $K$-levels toward a thermal distribution set by the vibrational temperature. Below $T_{\rm int}$, this coupling becomes inefficient and rotation effectively decouples from the internal degrees of freedom.

During thermal spikes in which the grain temperature exceeds $T_{\rm int}$, the $K$-levels are quickly thermalized to the vibrational temperature. As the grain cools below $T_{\rm int}$, the coupling shuts off and the $K$-distribution becomes frozen. Consequently, the $K$-state distribution of cold VSGs reflects the distribution at $T_{\rm int}$. Basing ourselves on \citet{draine:16}, we model $T_{\rm int}$ as a function of grain size and approximate it as
\begin{align}
\label{eq:internal_temperature}
T_{\mathrm{int}} (a_{\mathrm{eff}}) = 130 \ \mathrm{K} \left( \frac{a_{\rm eff}}{5 \ \rm \text{\AA}} \right)^{-\alpha},
\end{align}
with $\alpha=2.7$ for $a_{\rm eff}\leq5\ \text\AA$ and $\alpha=1.6$ for $a_{\rm eff}>5$ \AA. 

While Coriolis coupling efficiently redistributes the projection quantum number $K$ when $T_{\rm vib}>T_{\rm int}$, it conserves the total rotational quantum number $J$. Instead, the evolution of $J$ is governed by angular-momentum conservation under external torques \citep{draine:98}. Here, we assume that the ground state, $\gamma$, population of the rotational quantum states follows 
\begin{align}
\label{eq:rot_distr}
P_{\gamma JK} \propto 
\exp\!\left[-\frac{B J^2}{kT_{\rm rot}}\right]
\exp\!\left[-\frac{(A-B)K^2}{kT_{\rm int}}\right],
\end{align}
where the $K$-substructure is set by freeze-out at $T_{\rm int}$, while the total angular momentum distribution is characterized by the size-dependent rotational temperature $T_{\rm rot}$. In our simulations, we computed $T_{\rm rot}$ using the formalism of \citet{draine:98}, and adopting the prescribed torques of \citet{ali:09}. We note that detailed torque modeling shows that the resulting population distribution can deviate significantly from a single Boltzmann distribution \citep{ali:09,silsbee:11, zhang:25}. 

Unlike many treatments that assume the $K$-sublevels to be uniformly populated \citep{ali:09, silsbee:11, zhang:25}, we explicitly allow for population variations among $K$ at fixed $J$, since such imbalances are directly tied to grain alignment. We will show, however, that 
$K$-level imbalances are not a necessary condition for the emergence of polarized emission.
\subsection{Electric and magnetic dipole moment of a very small grain}
\subsubsection{Electric dipole moment}
Interstellar VSGs are expected to possess a permanent electric dipole moment.  The main contribution to the particle dipole moment arises from charge asymmetries within the grain due to electron displacements caused by differences in electronegativity among the constituent atoms. An additional contribution may arise from the net grain charge, although this term is often neglected \citep{ali:09}. The total electric dipole moment is commonly parameterized as scaling with the square root of the number of atoms in the grain,
$\mu_{\mathrm{tot}} = \beta \sqrt{N},
$
where $\beta \approx 0.4\,\mathrm{D}$ \citep{draine:98}. 

Beyond its magnitude, the orientation of the electric dipole moment is also important. For an axisymmetric VSG, it is convenient to decompose the electric dipole moment into components parallel and perpendicular to the symmetry axis, $\mu_{\parallel}$ and $\mu_{\perp}$. These are sometimes referred to as out-of-plane and in-plane components, respectively, particularly in the context of polycyclic aromatic hydrocarbons (PAHs). While the electric dipole orientation affects the emission spectrum, it is common to assume an approximately isotropic configuration, $\mu_{\parallel} = \frac{1}{2}\mu_{\perp}$ \citep{silsbee:11}.

Spinning dust emission arises from this permanent electric dipole moment as the VSG rotates. However, when radiative processes involve changes in the internal state of the grain, they are governed by the transition dipole moment between internal states. For a given transition between two internal states, $\gamma \to \gamma'$, we note the relevant electric dipole moment is $\braket{\gamma|\boldsymbol{\mu}|\gamma'} = \boldsymbol{\mu}^{\gamma \gamma'}$, which need not coincide with the permanent dipole moment of the grain. In fact, we should expect that electronic excitations can substantially modify both the magnitude and orientation of the grain electric dipole moment. In contrast, vibrational transition electric dipole moments within the electronic ground state are expected to remain close in magnitude and direction to the permanent dipole of the grain. In our simulations, we assume $\boldsymbol{\mu}^{\gamma \gamma'} \approx \boldsymbol{\mu}$.

In the remainder of this work, we consider three idealized dipole configurations: (i) a purely parallel dipole, $\mu_{\mathrm{tot}} = \mu_{\parallel}$; (ii) a purely perpendicular dipole, $\mu_{\mathrm{tot}} = \mu_{\perp}$; and (iii) a geometry-dependent dipole that follows the grain shape, defined by $\mu_{\perp}^2/\mu_{\parallel}^2 = 1/s$. 

\subsubsection{Magnetic dipole moment and precession}
Throughout this work, we assume that the ambient magnetic field defines the space-fixed symmetry axis of the VSG. This assumption is justified if the grain precesses around the magnetic field at a rate faster than any other interaction capable of reorienting the angular momentum, such as UV absorption or directional gas-grain collisions. The magnetic precession results from the torque $\boldsymbol{\tau} = \boldsymbol{m} \times \boldsymbol{B}$ acting on the grain magnetic moment, leading to Larmor precession of the angular momentum vector at a rate $\Omega_B \sim mB/J$ \citep{lazarian:07}.

The grain magnetic moment is commonly attributed to the Barnett effect, whereby rotation induces spin alignment that produces a magnetic moment \citep{purcell:79,andersson:15}. However, the establishment of spin alignment has to be achieved through some relaxation mechanism. In VSGs, relaxation pathways can be strongly suppressed in the cold phase \citep{draine:16}, while during thermal spikes angular momentum exchange between rotation and vibrational modes may disrupt the alignment of the spins. As a result, the Barnett effect may not be efficient for VSGs.

We therefore consider here a complementary magnetic moment. Even in the absence of electron or nuclear spins, a rotating grain acquires a magnetic moment purely from charge rotation \citep{flygare:71, lankhaar:18}. Because the spatial distributions of nuclear and electronic charge do not coincide exactly, their rotational currents do not cancel, generating a magnetic moment proportional to the angular momentum \citep{jackson:98}. In Appendix A.2, we derive that this rotational magnetic moment leads to a precession rate on the order of
$\Omega_B \sim 1.8 \times 10^{-5} \ \mathrm{s}^{-1}\  \left(\frac{a}{10\ \mathrm{\text{\AA}} } \right)^{-2} \left(\frac{B}{10\ \mathrm{\mu G} } \right).$
We note here that in contrast to the magnetic moment due to the Barnett effect, the rotational magnetic moment does not require any relaxation to emerge and provides therefore a robust alternative magnetic moment for VSGs. 

Under typical interstellar conditions, the precession rate due to the rotational magnetic moment exceeds other directional interaction rates, justifying our assumption that the magnetic field defines the space-fixed symmetry axis of the grain.

\section{Radiative transitions}
We now consider the interaction of VSGs with a radiation field. Since we cover interactions across many wavelength regimes, we must first verify the validity of the dipole approximation. This approximation allows us to describe the interaction of the radiation field with the VSG as a single coherent entity, rather than resolving local interactions at specific sites on the particle surface.

For the electric dipole approximation to hold, the size parameter
$x = \frac{2\pi a}{\lambda}
\approx 0.05 \left(\frac{a}{10\,\mathrm{\text{\AA}}}\right)\left(\frac{\lambda}{1216\,\mathrm{\text{\AA}}}\right)^{-1}
$
must remain small. To ensure that higher-order multipoles (in particular the electric quadrupole) are negligible, we adopt the conservative criterion $x<0.1$, which is satisfied for radii $a \lesssim 20\,\mathrm{\text{\AA}}$ at $\lambda \gtrsim 1216\,\mathrm{\text{\AA}}$. 

Having validated the electric-dipole approximation, we now examine the interaction between the radiation field and the VSG via its global electric dipole moment. We consider the transition between quantum states $\ket{(\gamma) J K M}$, and $\ket{(\gamma') J' K' M'}$. 
The radiative transition probability between the states 
$\ket{(\gamma)JKM}$ and $\ket{(\gamma')J'K'M'}$ in a directional radiation field depends explicitly on both
projection quantum numbers $K$ and $M$. Physically, this reflects that absorption
and emission depend on the orientation of the grain rotational angular momentum relative to
both the body-fixed symmetry axis (encoded by $K$) and the space-fixed axis set by
the radiation field (encoded by $M$). A complete expression is given in Eq.~(\ref{eq:dipole}) of the appendix.
Importantly, the angular dependence can be separated into body-fixed and space-fixed parts through
the Wigner--Eckart theorem \citep{zare:88}. We therefore postpone the explicitly directional (space-fixed)
aspects of the interaction and focus here on the body-fixed dependence encoded in $K$. The radiative transition
probability for a specific transition between $\ket{(\gamma)JK}$ and $\ket{(\gamma')J'K'}$, is a function of the Einstein $B$ coefficient 
$B_{\gamma JK \to \gamma' J' K'} = \frac{32 \pi^4}{3 h c} |\langle (\gamma K) J \| \mu \| (\gamma' K) J' \rangle|^2 
$, 
where the quantity in brackets is the so-called reduced matrix element of the dipole moment \citep{landi:84}.
Exploiting the separation between rotation and other interactions, we decouple the rotational state $(JK)$ from the internal structure ($\gamma$) within the reduced matrix element. In the limit of large rotational quantum numbers ($J \gg 1$), this separation yields the following expressions for the allowed transitions \citep{townes:55}
\begin{subequations}
\label{eq:einstein_symm}
\begin{align}
B_{\gamma JK \to \gamma' J\pm1 K'}  &= \frac{32 \pi^4}{3 h c}\left[ \frac{|\mu_{\parallel}^{\gamma \gamma'}|^2}{2} \left(1 - \left[\frac{K}{J}\right]^2 \right) \delta_{KK'} \right. \nonumber \\  &+ \left. 
\frac{|\mu_{\perp}^{\gamma \gamma'}|^2}{8} \left(1 \mp \frac{K}{J}\right)^2\delta_{KK'\pm1}\right]
\end{align}
for $P$- and $R$-branch transitions ($J \to J\pm 1$), and 
\begin{align}
B_{\gamma JK \to \gamma'J K'}  &= \frac{32 \pi^4}{3 h c}\left[ |\mu_{\parallel}^{\gamma \gamma'}|^2 \left[\frac{K}{J}\right]^2  \delta_{KK'} \right. \nonumber \\  &+ \left. \frac{|\mu_{\perp}^{\gamma \gamma'}|^2}{4} \left(1 - \left[\frac{K}{J}\right]^2 \right) \delta_{KK'\pm1} \right]
\end{align}
\end{subequations}
for $Q$-branch transitions ($J \to J$). Note the $\Delta K=0$ and $\Delta K=\pm1$ selection rules for transitions associated with the parallel and perpendicular components of the transition dipole moment \citep{townes:55}. 

For a specific transition $\ket{\gamma JK}\to \ket{\gamma' J' K'}$, the rate of absorption is given by
\begin{align}
\dot{N}_{\mathrm{abs}}^{\gamma JK\to \gamma' J' K'} = B_{\gamma JK \to \gamma'J K'} J_0^0 (\nu_{\gamma JK\to \gamma J' K'}),
\end{align}
which depends on the mean intensity (angularly integrated specific intensity) at the resonance frequency, $\nu_{\gamma JK\to \gamma' J' K'}$:
\begin{align}
J_0^0 (\nu_0) = \frac{1}{4\pi}\int d\hat{k} \int d\nu \ I_{\nu} (\hat{k}) \phi_{\nu-\nu_0},
\end{align}
where $\phi_{\nu-\nu_0}$ is a normalized line profile centered around the frequency $\nu_0$, and $\hat{k}$ indicates the radiation field direction. Since the rotational sublevels involved in an internal-state transition $\gamma \to \gamma'$ lie very close in energy, the specific intensity $J_0^0 (\nu)$ varies negligibly across the rotational band. It is therefore useful to define the total absorption rate from the initial, unaligned, state $\ket{\gamma JK}$ to all possible rotational states in the final electronic or vibrational manifold $\gamma'$:
\begin{align}
\dot{N}_{\mathrm{abs}}^{\gamma JK\to \gamma' } &= \sum_{J'K'}B_{\gamma JK \to \gamma'J K'} J_0^0 (\nu_{\gamma JK\to \gamma J' K'}) \nonumber \\
&\simeq J_0^0 (\nu_{\gamma JK\to \gamma'}) \sum_{J'K'}B_{\gamma JK \to \gamma' J' K'} 
\end{align}
where we have assumed the intensity as constant over the rotational manifold, centered around the frequency $\nu_{\gamma JK\to \gamma'}$. By summing over the allowed rotational transitions, we recognize from Eqs.~(\ref{eq:einstein_symm}) the total Einstein $B$-coefficient,
\begin{align}   
\label{eq:einstein_total} 
B_{\gamma JK \to \gamma'} =  \sum_{J'K'}B_{\gamma JK \to \gamma' J' K'}= \frac{32 \pi^4}{3 h c}\left[| \mu_{\parallel}^{\gamma \gamma'}|^2 + |\mu_{\perp}^{\gamma \gamma'} |^2 \right],
\end{align}
that depends only on the electric dipole moment's magnitude, and not its orientation.

\section{Alignment of very small dust grains}
Now we are going to consider the spatially angular aspects of the dipole interaction. 
We can recognize from Eq.~(\ref{eq:dipole}), that there is a clear dependence of the transition rate, $\ket{[\gamma K]JM} \to \ket{[\gamma' K']J'M'}$, for different $M \to M'$ transitions, depending on the direction and polarization of the radiation field. Consider the case that an ensemble of ground state VSGs are illuminated by an anisotropic ISRF, then degenerate $\ket{[\gamma K]JM}$ states, of different $M$, will be depopulated at different rates due to absorption events.

The most advantageous way to keep track of $M$-population imbalances is by representing the populations of particular states in terms of their irreducible tensor elements. We note the general irreducible tensor element \citep{landi:06}
\begin{subequations}
 \begin{align}
\rho_q^{k}([\gamma K] J) = \mathrm{tr} \left[ \hat{\rho}_q^k ([\gamma K]J) \hat{\rho} \right],
\end{align}
where $\hat{\rho}$ is the density operator. The irreducible tensor operator is defined as \citep{landi:06}
\begin{align}
\label{eq:irred}
\hat{\rho}^{k}_{q}([\gamma K] J)
&= \sum_{M M'} (-1)^{J-M}\,\sqrt{2k+1}
\begin{pmatrix}
J & J & k \\
M & -M' & q
\end{pmatrix} \nonumber \\ &\times
\ket{(\gamma) J K M}\bra{(\gamma) J K M'},
\end{align}
\end{subequations}
with $k$ and $q$ denoting the rank and projection of the operator and where the term in parentheses is a Wigner 3j-symbol. It can be immediately recognized that the irreducible element $\rho_0^0([\gamma K]J)$ is proportional to the total population of the $\ket{(\gamma) J K}$ energy level. We refer to this element as the isotropic population element. Conversely, in the limit of high $J$, and recalling that the $M$-projection refers to the space-fixed symmetry axis that is the magnetic field, it can be recognized that the irreducible tensor element, $\rho_0^{2}([\gamma K] J)$, is related to the classical alignment factor $R_{\boldsymbol{B}\boldsymbol{J}} = [3 \cos^2 \theta_{\boldsymbol{B} \boldsymbol{J}}-1]/2,$ used by \citet{draine:16}. The precise relation is $ \rho_0^{2}([\gamma K] J)/\rho_0^0([\gamma K]J) \to \sqrt{5} R_{\boldsymbol{B}\boldsymbol{J}}([\gamma K] J)$, where $R_{\boldsymbol{B}\boldsymbol{J}}([\gamma K] J)$ is the alignment factor of the particles in the state $\ket{[\gamma K]J}$. We refer to the irreducible tensor element $\rho_0^2([\gamma K]J)$ as the alignment population element.

Modeling the impact of a directional radiation field on the isotropic and alignment populations is a well studied problem in the theory of polarized line formation in stellar atmospheres and the interstellar medium \citep{landi:06, yan:06, goldreich:81, lankhaar:20a}. Now, we apply the concepts from this well established field of study to the alignment of VSGs. 

In the appendix, we derive the rate of change of the alignment population elements of a VSG quantum state exposed to a directional radiation field. In evaluating the evolution of a quantum level $\ket{[\gamma K]J}$, we assume that adjacent levels $\ket{[\gamma\, K\pm1]J\pm1}$ possess similar alignment properties, an approximation that is well justified in the large-$J$ regime that is appropriate for VSGs in interstellar conditions. Furthermore, when modeling the interaction with the ISRF, we assume that absorption proceeds between a single ground internal state $\gamma$ and one representative excited state $\gamma'$. This simplification neglects variations in transition dipole properties among the many possible absorption channels. We summarize the developments made in the appendix, and note the time evolution of the alignment population of a rotational level in the internal ground and excited states, $\gamma$ and $\gamma'$,
\begin{subequations}
\label{eq:two_level_rate}
\begin{align}
  \dot{\rho}^{2}_{0,g}
  &=
  \left[\dot{\rho}^{2}_{0,g}\right]_{\mathrm{abs}}^{\mathrm{dep}}
  + \Gamma\,\rho^{2}_{0,e}
  - A_0^2\,\rho^{2}_{0,g},
  \label{eq:rho2g_two_level_rate}\\[4pt]
  \dot{\rho}^{2}_{0,e}
  &=
  \left\langle\left[\dot{\rho}^{2}_{0,e}(K)\right]_{\mathrm{abs}}^{\mathrm{pump}}\right\rangle_{K}
  - \Gamma\,\rho^{2}_{0,e},
  \label{eq:rho2e_two_level_rate}
\end{align}
\end{subequations}
where we let $\rho^2_{0,g} \equiv \rho^{2}_{0}\!\left([\gamma K]J\right)$ and $\rho^2_{0,e} \equiv \rho^{2}_{0}\!\left([\gamma' K]J\right)$.
The time evolution of the alignment population of the ground and excited states are a function of interactions with the ISRF, that depopulates the ground state, captured in $ \left[\dot{\rho}^{2}_{0,g}\right]_{\mathrm{abs}}^{\mathrm{dep}}$, and populates the excited state, $\left[\dot{\rho}^{2}_{0,e}(K)\right]_{\mathrm{abs}}^{\mathrm{pump}}$. In the appendix, we derive this term to be
\begin{align}
\left[\dot{\rho}^{2}_{0,g}(K)\right]_{\mathrm{abs}}^{\mathrm{dep}} &\simeq - \dot{N}_{\mathrm{abs}} \left[-\sqrt{\frac{1}{5}} w \hat{\mu}_{\mathrm{anis}} P_2 (K/J)\ \rho_{0,g}^{0} \right. \nonumber \\   &+ \left. \left(1 - \frac{2}{7}w \hat{\mu}_{\mathrm{anis}} P_2 (K/J)\ \rho_{0,g}^2 \right) \right] \nonumber \\ 
&\simeq -\left[\dot{\rho}^{2}_{0,e}(K)\right]_{\mathrm{abs}}^{\mathrm{pump}} \nonumber
\end{align}
where the radiation anisotropy factor, $w$, and the dipole anisotropy parameter, $\hat{\mu}_{\mathrm{anis}}$ are defined later on in Eqs.~(\ref{eq:rad_anis}) and (\ref{eq:dip_anis}). Relaxation from the excited state, occurring at a rate $\Gamma$, depopulates the excited state level and repopulates the ground state level. The dominant mode of depolarization of the ground state level is through spontaneous emission, and indicated by $A_0^2$. Dominant depolarization through spontaneous emission is appropriate for the small VSGs that give rise to AME. This mode of depolarization is much slower than the relaxation rate, $\Gamma$, from the excited state, so we neglect it in the time evolution of the excited state. The quantity $P_2(K/J) = (3 (K/J)^2 - 1)/2$ is the second Legendre polynomial of the $K/J$ projection. Finally, we quote the time evolution of the excited state level in terms of $K$-averaged quantities, indicated by $\braket{}_K$, as Coriolis coupling, which occurs through open channels in the excited state, scrambles the $K$-levels at a much faster rate than the interactions detailed in the rate-equations.

With $\left[\dot{\rho}^{2}_{0,g}(K)\right]_{\mathrm{abs}}^{\mathrm{dep}}$ depending on the isotropic population $\rho^{0}_{0,g}$ while generating the alignment population component, $\rho^{2}_{0,g}$, absorption by an anisotropic radiation field acts as an alignment pump and therefore breaks isotropy. The angular dependence enters through the radiation and dipole anisotropy terms $w$, and $\hat{\mu}_{\mathrm{anis}}$. We note the dependence of the production of alignment on the body-fixed projection of the angular momentum through the term $P_2(K/J)$. Grains with $|K|/J\approx 1$ (rotation closely aligned with the symmetry axis) experience a (spatial) alignment drive orthogonal to that of grains with $|K|/J\ll 1$ (rotation largely perpendicular to the symmetry axis).

When discussing the (polarized) emissivity of the VSGs, it will become clear that the quantity of interest is the relative alignment of the quantum states, $\sigma_0^2([\gamma K] J) = \rho_0^2 ([\gamma K]J) / \rho_0^0([\gamma K]J)$, that we recall, is analogous to the alignment factor $R_{\boldsymbol{B}\boldsymbol{J}}$ used in the classical modeling of dust alignment. Assuming steady state for Eqs.~(\ref{eq:two_level_rate}), we recognize that the relative alignment can be solved for, for both the excited state as well as the ground state. In Eqs.~(\ref{eq:avg_dep_closed_form}) of the appendix, we quote the proper solution. For simplicity, we quote here the solution to leading order of $2/7 w \hat{\mu}_{\mathrm{anis}}P_2(K/J)$, 
\begin{align}
\sigma_0^2([\gamma K] J) &\simeq -\sqrt{\frac{1}{5}} \frac{w \hat{\mu}}{1 + \frac{A_0^2}{\dot{N}_{\mathrm{abs}}}} \left[P_2 (K/J) \right. \nonumber \\ &- \left. \frac{\Braket{P_2 (K/J) \rho_0^0([\gamma K]J)}_K}{\rho_0^0([\gamma K] J)} \right].
\end{align}
From this, we recognize that the relative alignment of the VSG quantum states, is a function of the following features.

First, the ratio between the rate of (aligning) directional absorption events and the rate of depolarization due to spontaneous emission events. In the appendix, we provide a rigorous formulation of the spontaneous emission depolarization rate; and detail how an order of magnitude estimate of this quantity is made around thermal frequency. Together with the estimate for the absorption rate (see Eq.~\ref{eq:abs_rate}), this leads to a ratio,
\begin{align}
\frac{A_0^2}{\dot{N}_{\mathrm{abs}}} &\approx \left(\frac{a}{3.9 \ \mathrm{\text{\AA}}}\right)^{-10}U^{-1} \left(\frac{T_{\mathrm{rot}}}{100 \ \mathrm{K}}\right) \nonumber \\ &\times \left(\frac{\beta}{0.3 \ \mathrm{D}} \right)^{2} \left(\frac{\rho}{3 \ \mathrm{g/cm^3}} \right)^{-2} 
\end{align}
that is a very strong function of the grain size. For standard parameters listed above, at the minimal grain size of $3.5\ \mathrm{\text{\AA}}$, the ratio $A_0^{2} / \dot{N}_{\mathrm{abs}} \approx 3.2$, indicating that alignment will be reduced by a factor $\sim 4$, over grains of $5 \ \mathrm{\text{\AA}}$, that have a ratio $A_0^{2} / \dot{N}_{\mathrm{abs}} \approx 0.09$. In fact, for grain sizes larger than $4.5\ \mathrm{\text{\AA}}$, depolarization is very slow compared to rate of (aligning) absorption interactions.

Second, the relative anisotropy of the ISRF, which is defined as
    \begin{align}
    \label{eq:rad_anis}
    w = \frac{\int d\hat{k}\ I_{\nu} (\Omega) P_{2} (\hat{k}\cdot \boldsymbol{B})}{\int d\hat{k}\ I_{\nu} (\Omega)}.
    \end{align}
    can vary from $1$ in case the ISRF is unidirectional and along the magnetic field direction to $-1/2$, in case the ISRF is unidirectional and oriented perpendicular to the magnetic field direction. The factor $w$ drops in magnitude when the radiation field gets more isotropic, and varies with the relation between the dominant ISRF direction and the magnetic field direction.

Third, the dipole anisotropy parameter, defined as
    \begin{align}
    \label{eq:dip_anis}
    \hat{\mu}_{\mathrm{anis}} = \frac{\mu_{\parallel}^2  - \frac{1}{2}\mu_{\perp}^2}{\mu_{\parallel}^2  + \mu_{\perp}^2},
    \end{align}
    which can range from $1$ for a purely parallel dipole moment, to $-1/2$ for a purely perpendicular dipole moment. 

Fourth, the quantum state body-fixed projection, $K$. States of $K /J\to  1$ are oppositely aligned to states of $K/J\to 0$. This means that they produce oppositely polarized radiation. However, we will see that averaged over the associated line strengths, the net polarization from an ensemble of VSGs will be nonzero.  

Fifth, the state alignment is a function of the distribution for the total population of $K$-levels for a particular angular momentum $J$. 

\begin{figure}
\centering
\includegraphics[width=0.9\linewidth]{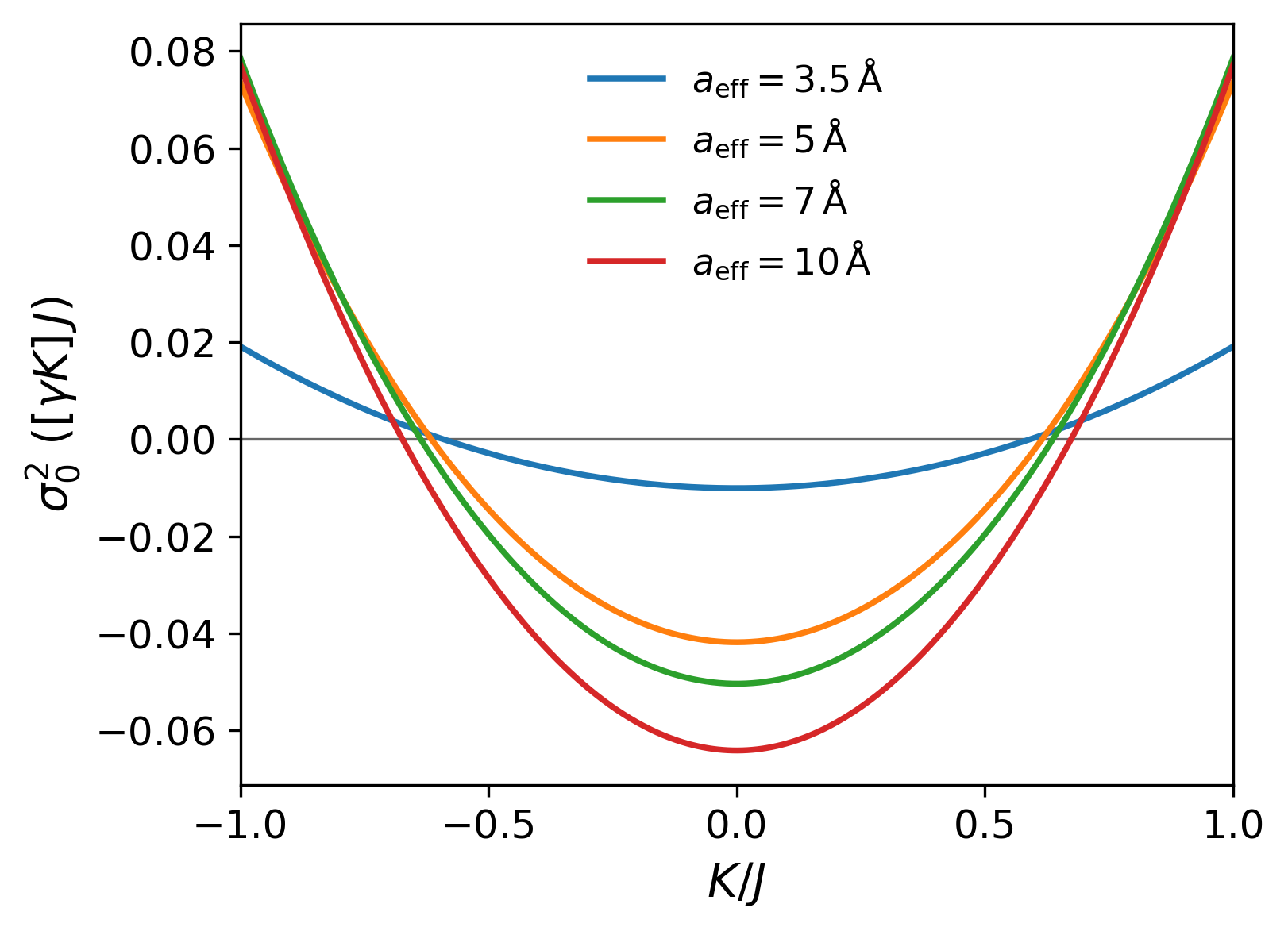}
\caption{Ground state relative alignment,
$\sigma^{2}_{0}\!\left([\gamma K]J\right)$, as a function of the normalized body-fixed projection $K/J$ for four effective grain sizes
$a_{\rm eff}=3.5,\,5,\,7,$ and $10\,\text{\AA}$. The curves are computed for $T_{\rm rot}=100\,\mathrm{K}$, $U=1$, $w=0.3$,
$\hat{\mu}_{\rm anis}=0.6$, and grain shape parameter $s=0.5$.}
\label{fig:ground_alignment}
\end{figure}

If all these parameters are known, then it is a trivial operation to compute the relative alignment of each quantum state of the VSG. In Fig.~\ref{fig:ground_alignment}, we show the relative alignment for a range of grain sizes, as a function of the body-fixed projection, $K/J$. The relative alignment increases strongly towards larger grains owing to their lower internal temperature and higher absorption rate relative to the depolarization rate. The size-dependent internal temperatures are
$T_{\rm int}=\{339.6,\,133.3,\,80.0,\,45.2\}\,\mathrm{K}$ and corresponding $\dot{N}_{\mathrm{abs}}/A_0^2=\{0.315,\,11.16,\,322.83,\,1.14\times10^{4}\}$, for the shown VSG sizes $a_{\rm eff} = \{3.5,\,5,\,7,\,10\}\,\mathrm{\text{\AA}}$.
The alignment is an even function of $K/J$ and changes sign near $\arccos (K/J) \approx 54.7^o$; the magic angle. 

As a final note, it is useful to point out that in the limit of evenly distributed $K$-levels, then the expression for the relative alignment simplifies to
\begin{align}
\label{eq:align_simplified_rotor}
\sigma_0^2([\gamma K] J) \approx -\sqrt{\frac{1}{5}} \frac{w \hat{\mu}_{\mathrm{anis}} P_2 (K/J) }{1 + \frac{A_0^2}{\dot{N}_{\mathrm{abs}}}}.
\end{align}
Noting the population in this form makes it particularly clear that ground-state VSGs are aligned at the particle level, but not at the population level ($\braket{\sigma_0^2(K/J)}_K\to 0)$. In fact, Eq.~(\ref{eq:rho2g_steady_balance}) shows that this holds generally, independent of the total population’s $K$-distribution. The reason is that anisotropic absorption in the dipole approximation produces no net torque. By symmetry, for illumination by a non-circularly polarized radiation field, the radiation-field tensor contains only rank-$0$ and rank-$2$ irreducible components, and therefore cannot exert a torque on the grain. In the next section we show, however, that because the emission strengths depend on $K/J$, the resulting emission need not be unpolarized even if the ground state is unaligned at the population level.

\section{The (polarized) emissivity of spinning dust}
Now, we turn to examine the effects of particle alignment on the polarization signature of the emission (and later also absorption) of VSGs. First, we set out to formulate a general form for the (polarized) emissivity from a single, aligned, VSG. At this point, we do not distinguish between vibrational (IR) emission, or purely rotational emission. For an aligned VSG of size $a$, the particle total intensity emissivity is
\begin{subequations}
\label{eq:emission_general}
\begin{align}
&\eta_I (a) = \sum_{\gamma J K,J' } P_{\gamma J K}(a) \left(1 + \frac{3 \cos^2 \vartheta - 1}{2 \sqrt{2}} w_{JJ'}^{(2)} \sigma_0^2 ([\gamma K] J) \right) \nonumber \\ &\times \sum_{\gamma' K'} \frac{h \nu_{\gamma J K \to \gamma' J' K'}}{4\pi} A_{\gamma J K \to \gamma' J' K'} \phi(\nu - \nu_{\gamma J K \to \gamma' J' K'}),
\end{align}
where we require that the states $\ket{\gamma JK}$ are higher in energy than the states $\ket{\gamma'J'K'}$, $A_{\gamma J K \to \gamma' J' K'}$ is the Einstein coefficient for the transition $\ket{\gamma JK} \to \ket{\gamma'J'K'}$ and $\nu_{\gamma J K \to \gamma' J' K'}$ is the transition frequency. The symbol $P_{\gamma J K} (a)$ denotes the population of the upper level $\ket{\gamma J K}$. We note that the population is normalized so that $\sum_{\gamma J K } P_{\gamma J K} (a) = 1$. The angle $\vartheta$, is the angle between the magnetic field direction and the emission direction. We indicated the line profile by $\phi(\nu - \nu_{\gamma J K \to \gamma' J' K'})$, and use the symbol, $w_{JJ'}^{(2)}$, for a geometric factor, introduced in \citep{landi:84}, that is specific to a $J\to J'$ transition. In the large-$J$ limit, this factor approaches,
\begin{align*}
w_{JJ'}^{(2)} &\simeq 
 \begin{cases}
    -\sqrt{\frac{2}{5}}, & J-J'=0 \\
    \sqrt{\frac{1}{10}}, & |J-J'|=1.
\end{cases}
\end{align*}
Proceeding to formulate the polarized emissivity, we note that we quantify the production of polarized specific intensity:
$Q = I_{\parallel} - I_{\perp},$
where $I_{\parallel}$ and $I_{\perp}$ are the intensities parallel and perpendicular to the magnetic field projection on the plane of the sky. In this frame, the polarized emissivity per particle is 
\begin{align}
&\eta_Q(a) = -\frac{3\sin^2 \vartheta}{2 \sqrt{2}}\sum_{\gamma J K,J' } P_{\gamma J K}(a) w_{JJ'}^{(2)} \sigma_0^2 ([\gamma K] J)\nonumber \\ &\times \sum_{\gamma' K'} \frac{h \nu_{\gamma J K \to \gamma' J' K'}}{4\pi} A_{\gamma J K \to \gamma' J' K'} \phi(\nu - \nu_{\gamma J K \to \gamma' J' K'}).
\end{align}
\end{subequations} 
\subsection{Pure rotational emissivity}
Now, we will specialize the emissivities to those occurring through purely rotational transitions. These are the emission modes that are hypothesized to give rise to AME \citep{draine:98, ali:09}. Pure rotational emission occurs within a fixed internal (vibrational/electronic) state, $\gamma\to\gamma$, and therefore corresponds to transitions entirely within a single rotational manifold. In the symmetric-top description that we employ, these transitions are grouped into ``branches'' labeled by the change in total rotational angular momentum $\Delta J$ and by whether the radiating dipole component is parallel or perpendicular to the grain symmetry axis. Below we list the allowed branches, together with their approximate line frequencies and Einstein $A$-coefficients in the large-$J$ limit.
\begin{subequations}
\paragraph{\textbf{\(P,\parallel\)}:} P-branch associated with the parallel dipole component, with
$\Delta J=-1$ and $\Delta K=0$. The emitted photon has frequency
$
\nu_{JK\to J-1,K}\approx 2BJ,
$
and the corresponding Einstein coefficient is
$A^{(P,\parallel)}_{JK\to J-1,K}
=\frac{64\pi^4 \nu^3}{3hc^3}\,\frac{|\mu_\parallel|^2}{2}\left(1-\frac{K^2}{J^2}\right).$
\paragraph{\textbf{\(P,\perp\)}:} P-branch associated with the perpendicular dipole component, with
$\Delta J=-1$ and $\Delta K=\pm 1$ transitions. The frequencies are
$
\nu_{JK\to J-1,K\pm1}\approx 2BJ \pm 2(B-A)K,$
and the Einstein coefficients are
$A^{(P,\perp)}_{JK\to J-1,K\pm1}
=\frac{64\pi^4 \nu^3}{3hc^3}\,\frac{|\mu_\perp|^2}{8}\left(1\mp \frac{K}{J}\right)^2.$
\paragraph{\textbf{\(Q,\perp\)}:} Q-branch associated with the perpendicular dipole component, with transitions with $\Delta J=0$ and $\Delta K=+1$ for $K\geq 0$ states, and $\Delta K=-1$ for $K\leq 0$ states. They are associated with frequencies
$
\nu_{JK\to J,K+1}\approx 2(B-A)K ,
$
and Einstein coefficients
$A^{(Q,\perp)}_{JK\to J,K+1}
=\frac{64\pi^4 \nu^3}{3hc^3}\,\frac{|\mu_\perp|^2}{2}\left(1-\frac{K^2}{J^2}\right).$
\end{subequations}
The total emissivity per grain is simply the sum of the different branch contributions,
\begin{align}
\label{eq:rot_tot}
\left[\eta_I (a)\right]_{\mathrm{rot}} = \left[\eta_I (a)\right]_{\mathrm{P,\parallel}} +\left[\eta_I (a)\right]_{\mathrm{P,\perp}} + \left[\eta_I(a)\right]_{\mathrm{Q,\perp}},
\end{align}
where,
\begin{subequations}
\label{eq:rot_tot_branch}
\begin{align}
&\left[ \eta_I (a)\right]_{\mathrm{P,\parallel}} = \frac{16 \pi^3}{3 c^3} \frac{|\mu_{\parallel}|^2}{2} \sum_{ JK} P_{\gamma JK} (a) \left(1 + \frac{3 \cos^2 \vartheta -1}{2\sqrt{20}}\sigma_0^2([\gamma K] J)  \right) \nonumber \\ &\times \left(1 - \frac{K^2}{J^2} \right) (2BJ)^4 \phi(\nu - 2BJ),
\\ 
&\left[ \eta_I (a)\right]_{\mathrm{P,\perp}} = \frac{16 \pi^3}{3 c^3} \frac{|\mu_{\perp}|^2}{4} \sum_{ JK} P_{\gamma JK} (a) \left(1 + \frac{3 \cos^2 \vartheta -1}{2\sqrt{20}}\sigma_0^2([\gamma K] J)  \right) \nonumber \\ &\times \left[ \left(1 - \frac{K}{J} \right)^2 \nu_+^4 \phi(\nu-\nu_+) + \left(1 + \frac{K}{J} \right)^2\nu_-^4 \phi(\nu-\nu_-) \right],
\\
&\left[ \eta_I (a)\right]_{\mathrm{Q,\perp}} = \frac{16 \pi^3}{3 c^3} \frac{|\mu_{\perp}|^2}{2} \sum_{ JK} P_{\gamma JK} (a) \left(1 - \frac{3 \cos^2 \vartheta -1}{2\sqrt{5}}\sigma_0^2([\gamma K] J)  \right) \nonumber \\ &\times \left(1 - \frac{K^2}{J^2} \right)
(2(B-A)K)^4 \phi(\nu - 2(B-A)|K|),
\end{align}
\end{subequations}
where we used the shorthand notation $\nu_{\pm} = 2BJ \pm 2(B-A)K$.
Proceeding to consider the polarized emissivity, per branch,
\begin{align}
\label{eq:rot_pol}
\left[\eta_Q(a)\right]_{\mathrm{rot}} = \left[\eta_Q(a)\right]_{\mathrm{P,\parallel}} +\left[\eta_Q(a)\right]_{\mathrm{P,\perp}} + \left[\eta_Q(a)\right]_{\mathrm{Q,\perp}},
\end{align}
we have,
\begin{subequations}
\label{eq:rot_pol_branch}
\begin{align}
&\left[ \eta_Q (a)\right]_{\mathrm{P,\parallel}} = -\frac{16 \pi^3}{3 c^3} \frac{3 \sin^2 \vartheta}{2\sqrt{20}} \frac{|\mu_{\parallel}|^2}{2} \sum_{ JK} P_{\gamma JK} (a) \sigma_0^2([\gamma K] J)  \nonumber \\ &\times \left(1 - \frac{K^2}{J^2} \right) (2BJ)^4 \phi(\nu - 2BJ),
\\ 
&\left[ \eta_Q (a)\right]_{\mathrm{P,\perp}} = -\frac{16 \pi^3}{3 c^3} \frac{3 \sin^2 \vartheta}{2\sqrt{20}} \frac{|\mu_{\perp}|^2}{4} \sum_{ JK} P_{\gamma JK} (a) \sigma_0^2([\gamma K] J)   \nonumber \\ &\times \left[ \left(1 - \frac{K}{J} \right)^2 \nu_+^4 \phi(\nu-\nu_+) + \left(1 + \frac{K}{J} \right)^2\nu_-^4 \phi(\nu-\nu_-) \right],
\\
&\left[ \eta_Q (a) \right]_{\mathrm{Q,\perp}} = \frac{16 \pi^3}{3 c^3} \frac{3 \sin^2 \vartheta}{2\sqrt{5}}\frac{|\mu_{\perp}|^2}{2} \sum_{ JK} P_{\gamma JK} (a) \sigma_0^2([\gamma K] J)  \nonumber \\ &\times \left(1 - \frac{K^2}{J^2} \right) (2(B-A)K)^4 \phi(\nu - 2(B-A)|K|).
\end{align}
\end{subequations}
Equations~(\ref{eq:rot_tot}-\ref{eq:rot_pol_branch}) can be solved in conjunction with Eq.~(\ref{eq:align_simplified_rotor}) to yield estimates for the total and polarized emissivities. However, as noted elsewhere, explicitly resolving the discrete rotational spectrum of VSGs may be unnecessary when modeling their emissivity, given the large angular momenta involved \citep{draine:98, ali:09}. Instead, Eqs.~(\ref{eq:rot_tot}-\ref{eq:rot_pol_branch}) may to good effect, be evaluated in the continuum approximation, where the line profile is replaced by a delta function $\phi(\nu) \to \delta (\nu)$, and the summations $\sum_{JK} \to \int_0^{\infty} dJ \int_{-J}^J dK$. We give the continuum expressions for the emissivities in the appendix. We have verified that the continuum expressions for the unpolarized and unaligned emissivity match the special case for $2\mu_{\parallel}^2 = \mu_{\perp}^2$, $B/A = 2$, and an even $K$-distribution, $P_{JK}(a) = P_J(a)$, that is derived classically in \citet{silsbee:11}.
\subsection{Analytical estimates for simplified dipole cases}
We derive analytically the polarization fraction from the emission from a single grain, which we assume has a size of $\sim 5$ \AA. We adopt Eq.~(\ref{eq:align_simplified_rotor}) to describe the alignment of this particle, $\sigma_0^2 ([\gamma K] J) \approx -\frac{1}{\sqrt{5}} w \hat{\mu}_{\mathrm{anis}} P_2 (K/J)$, where we have suppressed the depolarization factor $A_0^{2} / \dot{N}_{\mathrm{abs}}$ because it is very small at these sizes. As an initial estimate, we treat the cases where the dipole moment is purely parallel, and purely perpendicular. We assume the emission to be coming from close to $J=J_{\mathrm{th}}$. Furthermore, to derive the polarization fraction,
$p_Q = \frac{\eta_Q(a)}{\eta_I(a)},    
$
we drop factors $w_{JJ'} \sigma_0^2 ([\gamma K]J)$ beyond leading order. Under these approximations, the following expression for the polarization fraction of a parallel dipole moment is,
\begin{align}
p_Q^{\parallel} = -\frac{3 \sin^2 \vartheta}{2\sqrt{20}} \frac{\sum_K P_{\gamma J_{\mathrm{th}} K}(a)  \sigma_{0}^2 ([\gamma K] J) \left(1 - K^2/J^2\right)}{\sum_K n_{\gamma J_{\mathrm{th}} K} \left(1 - K^2/J^2\right)}.
\end{align}
Letting $P_{\gamma J_{\mathrm{th}} K}(a)$ be constant over $K$, using the continuum limit, $\sum_K \to \int_{-K}^K dK$, and using Eq.~(\ref{eq:align_simplified_rotor}), with $\hat{\mu}_{\mathrm{anis}}\to 1$ for a parallel dipole, we can derive
\begin{subequations}
\label{eq:rot_pol_approx}
\begin{align}
p_Q^{\parallel} = -\frac{3 \sin^2 \vartheta}{100} w = -3\%\ w \, \sin^2 \vartheta  .
\end{align}
Now we treat the case of a dipole moment perpendicular to the grain symmetry axis. For such a dipole moment, both Q-branch and P-branch transitions are allowed. In this analytical estimate, we only treat the P-branch transitions, since these are of highest frequency, and are thus strongly favored. We adopt the dipole anisotropy for a perpendicular dipole, $\hat{\mu}_{\mathrm{anis}}\to -1/2$, and adopt similar assumptions as before, to derive
\begin{align}
p_Q^{\perp} 
&= 1.5 \%\ \sin^2 \vartheta\ R\left(\delta\right) w,
\end{align}
\end{subequations}
where 
\[
R(\delta)=5\,\frac{7-56\delta+78\delta^{2}-48\delta^{3}+11\delta^{4}}{35-70\delta+84\delta^{2}-42\delta^{3}+9\delta^{4}}\,,
\]
is a function of $\delta = (B-A)/B$, that models the dependence of the polarization on particle non-sphericity. $R(\delta)$ equals $1$, when $A \approx B$, while it equals $-2.5$ when $B \gg A$. We find that for $s=1/2$,  $p_Q^{\perp} = -1.1 \%\ \sin^2 \vartheta\ w$.

These analytical estimates illustrate several generic features. First, the predicted polarization reverses sign between the purely parallel and purely perpendicular dipole cases. This sign flip is controlled by the dipole-anisotropy parameter $\hat{\mu}_{\mathrm{anis}}$ in Eq.~(\ref{eq:align_simplified_rotor}). Second, even under the assumption of an even $K$-distribution, we obtain nonzero polarization fractions. Polarization emerges despite alignment on a population-level averaging out to zero. The polarization dependence survives because the emissivities are strong functions of the $K/J$-state. 

Importantly, the resulting linear polarization is oriented either parallel or perpendicular to the projection of the magnetic field on the plane of the sky. There is a $90^\circ$ ambiguity analogous to that encountered in the Goldreich-Kylafis effect and in maser polarization \citep{goldreich:81, lankhaar:20a, lankhaar:24a}. Here, the polarization orientation is set by the ISRF anisotropy, $w$, and by dipole-anisotropy, $\hat{\mu}_{\mathrm{anis}}$, both of which can be positive or negative. Without independent constraints on these two quantities, the polarization direction cannot be uniquely related to the magnetic field direction.
\subsection{Simulations}

\begin{figure}[t]
    \centering
    \includegraphics[width=\linewidth]{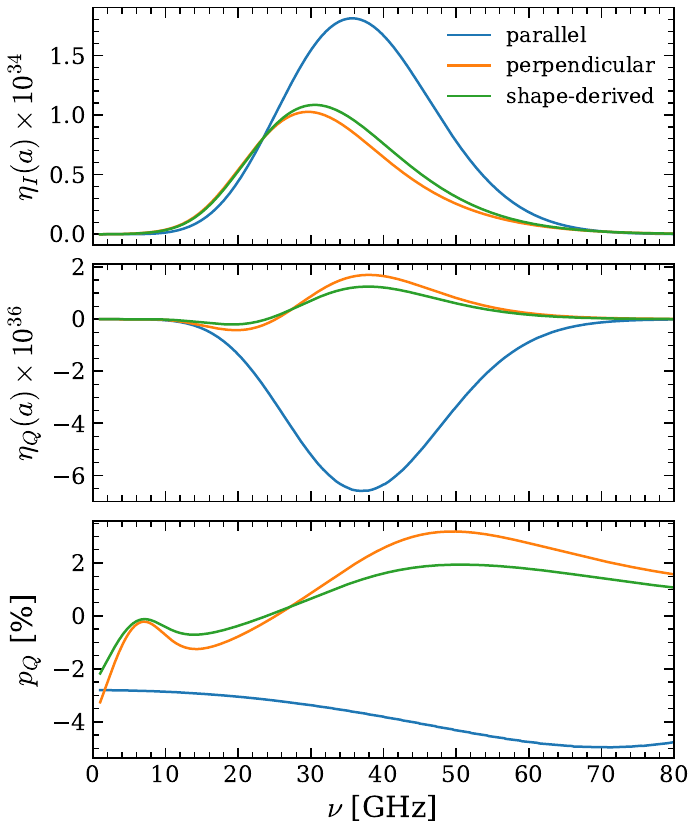}
    \caption{(Upper) emission, (middle) polarized emission, and (lower) polarization fraction of an oblate ($s=1/2$) dust-like particle of $a_{\rm eff}=5$~\AA. A radiation field anisotropy of $w=1$ is assumed. Inside the figures, the cases of a fully parallel and perpendicularly aligned dipole moment are shown, as well as the case of the shape-derived dipole moment. Units of $\eta_{I,Q} (a)$ are $\mathrm{erg\ s^{-1}\ Hz^{-1}\ sr^{-1}}$.}
    \label{fig:overview_a5}
\end{figure}

Figure~\ref{fig:overview_a5} summarizes the simulated polarization signatures for a single oblate, dust-like VSG of $5\,\mathrm{\text{\AA}}$. We used the continuum approximation expressions for the emissivities (Appendix E), in conjunction with Eq.~(\ref{eq:avg_dep_closed_form}) to generate the plots. We assumed that the rotational population of our $5$ \AA grain follow a Boltzmann distribution at a rotational temperature $T_{\mathrm{rot}}=100\ \mathrm{K}$, which sets the occupation of the $J$ ladder, while an internal temperature $T_{\mathrm{int}}$ determines the distribution over $K$ sublevels within a given $J$ manifold. 
We set the radiation anisotropy of the simulations $w=1$. We modeled the idealized cases where the dipole moment of the grain is along and perpendicular to the symmetry axis, as well as a shape-derived dipole moment. 

Before discussing the polarization in detail, it is useful to interpret the total emissivity profiles in Fig.~\ref{fig:overview_a5}. The total emission spectra for purely parallel and purely perpendicular dipole moments differ in that the perpendicular dipole spectrum is weaker by about $60\%$. Furthermore, the perpendicular-dipole case exhibits a peak at a somewhat lower frequency and a slightly more asymmetric profile.

Now we proceed to discuss the polarization signatures. For a dipole moment parallel to the symmetry axis, the polarization fraction is found to be negative over the full frequency range. The polarized emissivity peaks at somewhat higher frequencies than the total emissivity, which results in a rapidly increasing polarization fraction on the high-frequency side of the total-emission peak. In this regime, polarization fractions $<-3\%$ are reached close to the frequency where the emission is maximal. For a dipole moment perpendicular to the symmetry axis, the polarization is negative at low frequencies, but flips sign at higher frequencies. This behavior reflects the dominance of low-$K$ transitions at the low-frequency end, while high $K$ transitions dominate at higher frequencies. As in the parallel-dipole case, the polarized emission peak is shifted to higher frequencies relative to the peak of the total emission. Near the (positive) maximum of the polarized emissivity, polarization fractions of $\sim 2$--$3\%$ are predicted. The (polarized) emissivity profiles of the shape-derived dipole moment is broadly similar to the perpendicular dipole case, but we note that generally, the polarization fraction is slightly weaker. 

\begin{figure}[t]
    \centering
    \includegraphics[width=\linewidth]{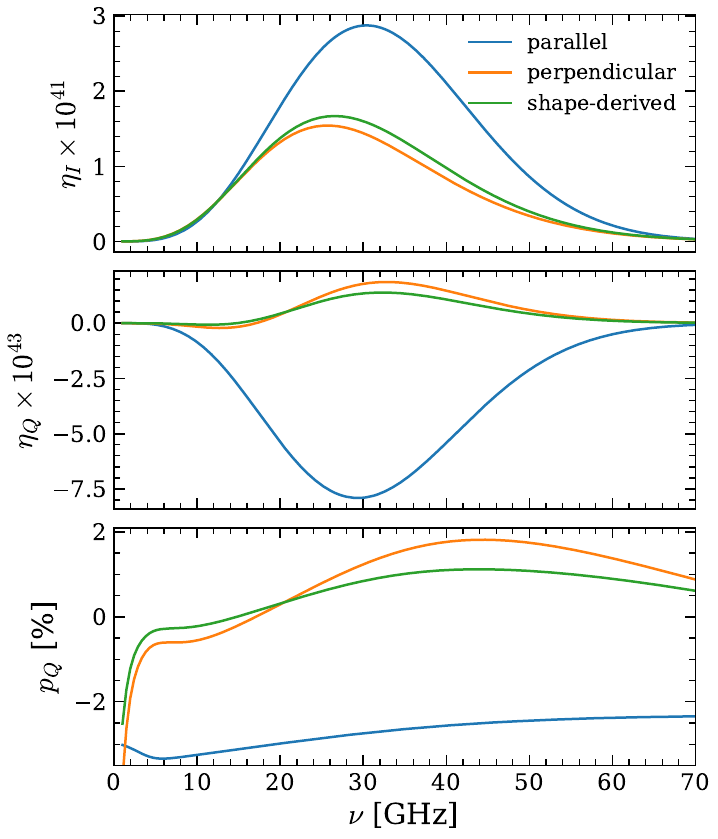}
    \caption{(Upper) emission, (middle) polarized emission, and (lower) polarization fraction of oblate ($s=1/2$) dust-like particles of grain size distribution of Eq.~(\ref{eq:size_distribution}). A radiation field anisotropy of $w=1$ is assumed. Inside the figures, the cases of a fully parallel and perpendicularly aligned dipole moment are shown, as well as the case of the shape-derived dipole moment. Units of $\eta_{I,Q}$ are $\mathrm{erg\ s^{-1}\ Hz^{-1}\ sr^{-1}\ H^{-1}}$.}
    \label{fig:size_distribution}
\end{figure}

In Fig.~\ref{fig:size_distribution}, the emissivity of a size-distributed sample of VSGs is given. The adopted size-distribution is 
\begin{align}
\label{eq:size_distribution}
\frac{1}{n_{\rm H}}\,\frac{dn}{d\log a_{\rm eff}} = 1.3\times 10^{-6}\,\exp\!\left[-\frac{\log^{2}\!\left(a_{\rm eff}/(3.5\,\mathrm{\text{\AA}})\right)}{0.32}\right]\,\mathrm{H}^{-1}.
\end{align}
Rotational temperatures were derived by balancing grain torques (see Section 2.2), and internal temperatures of Eq.~(\ref{eq:internal_temperature}) were used. We plot the total and polarized emission for a grain-size distribution assuming parallel, perpendicular, and shape-derived dipole moments. The total-intensity spectra resemble the features of the single-grain profiles, where perpendicular dipole emission branches are shifted to slightly lower frequencies, while exhibiting a more asymmetric profile. The emission peaks near $\sim 30\,\mathrm{GHz}$.

The polarization spectra show the same generic behavior as for a single grain: the polarized-emission peak is shifted to higher frequencies relative to the total-emission peak. The resulting fractional polarizations are smaller, compared to the $5\ \mathrm{\text{\AA}}$ case, because the smallest grains are a significant component of the population, while also being associated with more efficient depolarization. At the frequency of maximal emission we find polarization fractions of order $\sim -2\%$ for the parallel-dipole case and $\sim 0.75$--$1.5\%$ for the perpendicular and shape-derived cases.

\section{UV and IR polarization signatures}
\subsection{Absorption to excited electronic or vibrational states}
We now proceed to consider the (polarized) absorption signatures from VSGs. Absorption will be considered at IR wavelengths or shorter, where stimulated emission events can be neglected. Here, the single-grain absorption coefficient for the total intensity is given by
\begin{subequations}
\label{eq:abs_nostim}
\begin{align}
&\kappa_I(a) = \sum_{\gamma J K, J'} P_{\gamma JK}(a) \left(1 + \frac{3 \cos^2 \vartheta - 1}{2 \sqrt{2}} w_{JJ'}^{(2)} \sigma_0^2 ([\gamma K] J) \right) \nonumber \\ &\times \sum_{\gamma' K'} \frac{h \nu_{\gamma J K \to \gamma' J' K'}}{4\pi} B_{\gamma J K \to \gamma' J' K'} \phi(\nu - \nu_{\gamma J K \to \gamma' J' K'}),
\end{align}
where we require the state $\ket{\gamma' J' K'}$ to be of higher energy than $\ket{\gamma J K}$. The polarized single-grain absorption coefficient is
\begin{align}
&\kappa_Q (a)= -\frac{3 \sin^2 \vartheta}{2 \sqrt{2}} \sum_{\gamma J K, J'} P_{\gamma JK}(a)  w_{JJ'}^{(2)} \sigma_0^2 ([\gamma K] J)  \nonumber \\ &\times \sum_{\gamma' K'} \frac{h \nu_{\gamma J K \to \gamma' J' K'}}{4\pi} B_{\gamma J K \to \gamma' J' K'} \phi(\nu - \nu_{\gamma J K \to \gamma' J' K'}).
\end{align}
\end{subequations}
Next, we make the assumption that frequency modulations due to the rotational structure are negligible in these absorption events. Furthermore, we adopt the high-$J$ approximation, and obtain
\begin{subequations}
\begin{align}
&\kappa_I (a)= \sum_{\gamma J K} P_{\gamma JK}(a)  \left(1 - \frac{3 \cos^2 \vartheta - 1}{2 \sqrt{5}} \hat{\mu}_{\mathrm{anis}} \sigma_0^2 ([\gamma K] J) \right. \nonumber \\ &\times \left.  P_{2}(K/J) \right) \sum_{\gamma'} \frac{h \nu_{\gamma J K \to \gamma'}}{4\pi} B_{\gamma J K \to \gamma'} \phi(\nu - \nu_{\gamma J K \to \gamma'}),\\
&\kappa_Q (a) = \frac{3 \sin^2 \vartheta}{2 \sqrt{5}} \hat{\mu}_{\mathrm{anis}} \sum_{\gamma J K} P_{\gamma JK}(a) \sigma_0^2 ([\gamma K] J) P_{2}(K/J) \nonumber \\ &\times \sum_{\gamma'} \frac{h \nu_{\gamma\to \gamma'}}{4\pi} B_{\gamma J K \to \gamma'} \phi(\nu - \nu_{\gamma \to \gamma'}),
\end{align}
\end{subequations}
The alignment populations have a strong relation to the relevant $K$-states, with $\sigma([\gamma K J]) \propto P_2(K/J)$. Thus, the summation over the $K$-states that is implicit in the expression for the polarized absorption coefficient, does not tend to zero as it did for the emission expression; also under the assumption of a constant population within the $K$ ladder.
\subsection{Analytical estimates for polarization fractions due to IR/VIS/UV absorption}
We proceed to consider the polarization in absorption from a typical $5$ \AA\, dust grain. Specializing our definition of the polarization fraction to an absorption line
\begin{align}
p_Q &= \frac{I_{\parallel} - I_{\perp}}{I_{\parallel} + I_{\perp}} = \frac{e^{-(\kappa_I+\kappa_Q)s} - e^{-(\kappa_I-\kappa_Q)s}}{e^{-(\kappa_I+\kappa_Q)s} + e^{-(\kappa_I-\kappa_Q)s}} \nonumber \\ 
&= \tanh \kappa_Q s \simeq \kappa_Q s = \frac{\kappa_Q}{\kappa_I} \tau_I,
\end{align}
we note that there is a definite dependence of the polarization fraction on the line optical depth, $\tau_I$.

We now evaluate the fraction $\kappa_Q / \kappa_I$, and assume the level alignment follows Eq.~(\ref{eq:align_simplified_rotor}). We neglect any terms $w_{JJ'}^2 \sigma_0^2 ([\gamma K]J)$ that are beyond leading order, and assume the vibronic transition of interest, is between two single electronic-vibrational states, $\gamma \to \gamma'$. We then can note the resulting fraction
\begin{align}
\frac{\kappa_Q}{\kappa_I} \approx -\frac{3 \sin^2 \vartheta}{10} \hat{\mu}_{\mathrm{anis}}^2 w \frac{\sum_{ JK} P_{\gamma JK} (a) P_2(K/J)^2}{\sum_{JK} P_{\gamma JK} (a)}. 
\end{align}
We assume the particle is populated around $J_{\mathrm{th}}$, and adopt the continuum approximation for the $K$-state distribution. Then, we can evaluate
\begin{align}
\frac{\kappa_Q}{\kappa_I} \approx -\frac{3 }{25} \sin^2 \vartheta \hat{\mu}_{\mathrm{anis}}^2 w \approx 12\% \ \sin^2 \vartheta\ \hat{\mu}_{\mathrm{anis}}^2 w .
\end{align}
Contrary to the rotational emission, the configuration of the dipole moment with respect to the particle symmetry axis does not impact the relation of the magnetic field direction to the polarization direction. However, we recall that the ISRF anisotropy parameter, $w$, can assume both positive or negative values, depending on its orientation with respect to the magnetic field. Under optimal conditions, when $w \to 1$ and $\vartheta = 90^o$, $p_Q^{\parallel} = 12\% \ \tau_I$ and $p_Q^{\perp} = 3\% \ \tau_I$
\subsection{Polarization and IR emission}
We now consider the polarized emissivity from a thermally excited grain. The population averages we derived in Section 4 may not be representative for such transitions, as they likely occur on a time scale that is shorter than the magnetic precession time, after the ISRF excitation of the grain. Therefore, the averaging over magnetic precession that is assumed in the developments of Section 4, is not appropriate. The picture of alignment of VSGs that give rise to IR emission is more dynamical. VSGs that are in the ground state, are aligned with respect to the magnetic field. After the excitation, an alignment is induced that is with respect to the exciting photon direction (and polarization). As such, alignment will be produced that is with respect to an intermediate direction between magnetic field and the photon direction. Magnetic precession will force the VSG alignment direction (slowly) to the magnetic field direction, but this will be completed only after some complete precession rotations. Any polarization produced by the excited VSG will be with respect to the intermediate alignment direction after the excitation. 

It is perhaps still instructive to consider the idealized case where the direction of the exciting photon coincides with the magnetic field direction. In such a case, we can formulate the polarized emissivity cleanly. We assume the grain is in the excited state $\ket{[\gamma' K']J'}$, and we consider the vibrational or electronic relaxation, through a spontaneous emission event. Assuming $\nu_{\gamma' J' K' \to \gamma J K} \approx \nu_{\gamma' \to \gamma}$, we can note the per-grain polarized emissivity, to all possible rotational levels in a lower vibronic state, $\gamma$, as
\begin{align}
&\eta_Q(a) = \frac{3 \sin^2 \vartheta}{2 \sqrt{5}} \frac{h \nu_{\gamma\to \gamma'}}{4\pi}  \sum_{J'}A_{\gamma' J' K' \to \gamma} \hat{\mu}_{\mathrm{anis}} \nonumber \\ & \times \Braket{\sigma_0^2 ([\gamma' K'] J')}_{K'}\phi(\nu - \nu_{\gamma \to \gamma'})\sum_{K'} P_{\gamma' J'K'}(a)  P_{2}(K'/J') .
\end{align}
Owing to rapid equilibration of the $K$-states with the internal vibrational modes, the relevant alignment is the one averaged over all accessible $K'$-states. In the case of a high vibrational temperature, $T_{\rm vib} \gg T_{\rm rot}$, the $K'$-states are effectively randomized, such that $P_{\gamma' J'K'}(a) = P_{\gamma' J'}(a)$, and thus the final summation over $K'$ vanishes. For such grain populations the emission is expected to be unpolarized, even for grain populations that are aligned on average.

\section{Discussion}

\subsection{Anisotropic illumination alignment mechanism}
For larger interstellar grains, the dominant alignment paradigm is radiative-torque alignment (RAT): anisotropic starlight exerts systematic torques on helical grains, which spins them up to suprathermal rotation. Various relaxation processes subsequently lead to alignment of the rotational angular momentum vector along the magnetic field \citep{lazarian:07b, andersson:15}. For VSGs, however, RATs become inefficient because radiative torques drop rapidly in the Rayleigh regime \citep{lazarian:07}. Consequently, VSGs are not spun up and efficiently aligned by RATs.

A natural alternative is paramagnetic alignment via Davis--Greenstein (DG) relaxation. For a rotating paramagnetic grain, the component of the ambient interstellar magnetic field perpendicular to the spin axis appears time-dependent in the grain frame, inducing a lagging magnetization. The associated dissipation produces a torque that drives the grain angular momentum $\boldsymbol{J}$ toward alignment with the external magnetic field $\boldsymbol{B}$ \citep{davis:51}. \citet{lazarian:00} showed that this dissipation is very efficient in VSGs through certain resonances; and dubbed this mechanism resonance paramagnetic relaxation. This mechanism has been considered a promising route to partial alignment of ultrasmall grains and hence a potentially polarized spinning-dust signal \citep{hoang:13, hoang:16b}.

More recently, \citet{draine:16} emphasized that DG-/resonance-type alignment of VSGs implies the requirement that heat is efficiently dissipated into vibrational modes. For grains that are very small, and moreover cold most of the time, the vibrational density of states is sparse and such dissipation channels are effectively closed. Because of the closed dissipation channels, the paramagnetic alignment of grains is suppressed heavily \citep{draine:16}. In the size range most relevant for AME ($a_{\rm eff} \lesssim 7\,\mathrm{\text{\AA}}$), the predicted alignment, and subsequent linear polarization of rotational emission, becomes extremely small, at the level of $\lesssim 10^{-6}$ \citep{draine:16}.

We consider a different mechanism for aligning VSGs and polarizing their emission: interaction with an anisotropic radiation field. Our polarization model is inspired by treatments of line polarization in stellar atmospheres and the interstellar medium \citep{landi:06, goldreich:81, yan:06, lankhaar:20a}. The formalism is fully quantum mechanical, but exploits simplified limits appropriate for the large rotational angular momenta exhibited by VSGs.

In our model, VSG alignment arises because absorption from an anisotropic ISRF depends on the grain's angular momentum vector orientation relative to the magnetic field. We consider a simplified but representative system consisting of two vibronic manifolds, each carrying a ladder of rotational states. The time evolution of the rotational populations in the ground and excited manifolds is followed to obtain the steady-state distribution under anisotropic illumination. The absorption rate depends on both the projection of the angular momentum onto the body-fixed grain axis,
$
K/J = \hat{\boldsymbol{J}}\!\cdot\!\hat{\boldsymbol{a}}_1,
$
and onto the magnetic-field direction,
$
M/J = \hat{\boldsymbol{J}}\!\cdot\!\hat{\boldsymbol{B}}.
$
For positive (negative) $\mu_{\rm anis}\,w$ (dipole anisotropy times radiation-field anisotropy), states with small (large) $K/J$ are preferentially depleted at small (large) $M/J$ while they have a longer lifetime at large (small) $M/J$. Since the lifetime of a VSG becomes a function of its $M$-projection, anisotropic illumination will lead to their partial alignment. Importantly, even though VSGs will be aligned on an individual particle level, they are not aligned on the population level. Nevertheless, because radiative emission propensities depend on $K$, a population of individually aligned grains can still produce polarized emission even when the ensemble exhibits little or no net alignment.

We constructed our model to be general to any VSG that may be considered in the dipole approximation, and can be effectively represented as an oblate rotor at large rotational angular momentum (see sections $2$ and $3$). In other words, the model we present works for nanosilicates and PAHs alike. In the context of the alignment of PAHs, we note the related work of \citet{hoang:18}, that considered anisotropic illumination as a way to enhance alignment in PAHs that proceeds primarily through resonance paramagnetic relaxation. Their model of anisotropic absorption and infrared emission was classical, and it is difficult to compare against the first-principle quantum mechanical formalism of radiative interactions that we present here. At any rate, \citet{hoang:18} showed that the inclusion of resonance relaxation channels produced high degrees of alignment, in particular towards reflection nebulae. Polarization fractions from our modeling are more modest, but do not require the controversial mechanism of resonance paramagnetic relaxation \citep{draine:16}. The different observational predictions between the models should be confronted by future observations. 

Finally, our model highlights the dependence of particle alignment on the body-fixed projection, $K$, of the grain particle; and its manifestation in the polarization signal due to $K$-dependent emission probability. The body-fixed angular momentum projection is not a quantity that current classical Langevin/torque based modeling of the alignment includes \citep{draine:16, hoang:18}. We recommend that future alignment models explicitly incorporate body-fixed projection–resolved dynamics.

\subsection{Observational constraints and perspective}
\label{sec:observations}
Assuming AME arises from spinning-dust emission, our framework predicts linear polarization oriented either parallel or perpendicular to the plane-of-sky magnetic field projection. The polarization fraction depends on the product of the anisotropy parameter $w$ and the effective dipole anisotropy $\hat{\mu}_{\mathrm{anis}}$. For realistic dipole models and maximally anisotropic radiation fields---representative of strongly illuminated environments such as PDRs or reflection nebulae---we obtain polarization fractions of order $1$–$3\%$ at AME frequencies. In more typical CNM conditions, where the ISRF anisotropy is small (Eq.~\ref{eq:rad_anis}), the polarization fraction is correspondingly reduced, likely by an order of magnitude.

A key feature of this framework is that it provides self-consistent predictions for both microwave rotational emission and complementary UV/IR absorption polarization diagnostics, while predicting negligible to very low polarization in IR emission bands. 

\subsubsection{Microwave polarization of AME.}
AME is generally weakly polarized. Strongest polarization is found towards the Perseus molecular complex (G159.6$-$18.5), where \citet{battistelli:06} reported a total polarization fraction $\approx 3.4\%$ at 11 GHz. This result remains unusual in that subsequent measurements at higher frequencies have typically been consistent with zero polarization or have yielded upper limits, including limits derived from WMAP-band analyses toward Perseus and $\rho$~Ophiuchi (e.g.\ \citealt{dickinson:11,lopezcaraballo:11}). For the dark cloud Lynds~1622, \citet{mason:09} reported an upper limit of a few per cent at $\sim$10 GHz. For the H{\sc ii} region LPH96, CBI observations yielded no detection with an upper limit at the percent level \citep{dickinson:06}. \citet{battistelli:15} report a $(2.2\pm 0.5)\%$ polarized signal towards the H{\sc ii} region RCW175, but this polarization can plausibly be attributed to a sub-dominant synchrotron component rather than to AME itself. The most stringent constraints to date come from QUIJOTE and related analyses toward bright Galactic regions. In particular, \citet{genovasantos:17} obtained stringent upper limits toward W43, of $<0.39\%$ at 16.7 GHz, $<0.52\%$ at 22.7 GHz, and $<0.22\%$ at 40.6 GHz. \citet{gonzalez:25} further tightened these constraints and found sub percent polarization towards $\rho$~Ophiuchi and Perseus, too.

These constraints are compatible with a scenario in which the effective anisotropy $w$ is modest in the relevant emitting volumes and/or the population-averaged dipole anisotropy is small. The apparent low-frequency Perseus result \citep{battistelli:06} is therefore of particular interest: if confirmed, it would point either to unusually favorable anisotropy/alignment conditions in Perseus or to an additional emission mechanism contributing to the AME band in that region. Deeper polarization observations of Perseus with frequency coverage spanning the AME peak would be especially valuable.

\subsubsection{Constraints from UV/optical/IR spectropolarimetry.}
PAHs are often invoked as plausible carriers of the $2175\ \mathrm{\text{\AA}}$ bump, and UV spectropolarimetry has been used to test whether that carrier population is aligned. Early measurements established UV spectropolarimetry across the bump and searched for a polarized ``bump'' component \citep{clayton:92,wolff:97,martin:99}. In specific sources, there is evidence of a change in polarization behavior around the $2175\ \mathrm{\text{\AA}}$ feature, but the general outcome is that the bump is typically weakly polarized compared to what would be expected if its carrier were strongly aligned in the same manner as larger grains \citep{martin:99}. However, even if PAHs contribute to the bump, it remains unclear whether the PAH population responsible is the same as (or representative of) the smallest grains implicated in AME \citep{draine:03, hensley:16, sponseller:25}.


More broadly, beyond the behavior specifically associated with the 2175 Å bump, it is also worth noting that some sightlines show UV polarization exceeding the standard Serkowski extrapolation, which has been interpreted as possible evidence that alignment extends to VSGs in particular environments \citep{clayton:95}. Such an observation is qualitatively consistent with our mechanism, where VSG alignment properties vary strongly with environment, and are enhanced in regions with highly anisotropic radiation fields.

IR spectropolarimetry also provides a test of whether candidate small-grain carriers (including PAH-related populations) are aligned. In the diffuse ISM, spectropolarimetry of the 3.4 $\mu$m aliphatic C--H stretch absorption feature has yielded non-detections at the $\lesssim$0.1\% level in representative Galactic-center sightlines \citep{chiar:06}, which has been widely interpreted as evidence that the 3.4 $\mu$m feature carriers are not co-aligned with the large, efficiently polarizing silicate population. For PAH-related emission features, early searches placed upper limits at the percent level for the 3.3 and 11.3 $\mu$m bands in reflection nebulae, with one tentative detection \citep{sellgren:88}. Spectropolarimetry of the active galaxies NGC~1068 and NGC~4151 likewise showed the 11.3 $\mu$m PAH feature to be unpolarized \citep{lopez-rodriguez:16, lopez-rodriguez:18}. \citet{zhang:17b} has reported the 11.3 $\mu$m feature to be polarized at a level of $1.9\%$ towards the MWC 1080 nebula, but an independent reanalysis of the same spectropolarimetric data by \citet{lopez-rodriguez:26} rebutted this detection claim and instead found the data to be consistent with the instrumental polarization. To date, there are therefore no confirmed detections of polarized PAH emission features. This is in line with our prediction of negligible polarization in IR emission from VSGs aligned by anisotropic radiation. 

\subsubsection{Joint polarization diagnostics}
A clear prediction of our model is the tight association of polarization fraction with the anisotropy of the ISRF radiation field. In environments with large $|w|$ (strongly illuminated PDRs / reflection nebulae), we expect both AME rotational emission and UV/optical/IR absorption features to be polarized, with their polarization fractions governed by the same anisotropic-illumination alignment physics. In contrast, IR emission features are predicted to be negligibly polarized, even when the grains are aligned, owing to K-scrambling during thermal spikes. Joint measurements of (i) AME polarization, (ii) UV bump/feature absorption polarization, and (iii) PAH-band emission polarization in such environments therefore offer a clean path to falsify or validate the predictions of our model. 
    
\subsection{Some remarks on the total emissivity of spinning dust}
Expressions for the rotational emissivity of non-spherical, axisymmetric grains have been developed previously and are implemented in \textsc{spdust} \citep{ali:09, silsbee:11}, which remains a widely used reference for modeling AME from spinning dust. In the published \textsc{spdust} formulations, however, the emissivity expressions are typically presented for a restricted set of assumptions: homogeneous population of $K$-sublevels within each $J$-manifold, a particular dipole-moment parametrization (often the ``isotropic'' case with a fixed ratio between in-plane and out-of-plane components), and a specific choice of grain oblateness. Our quantum-mechanical treatment reproduces these classical limits, but is considerably more general: we derive emissivities that remain valid for arbitrary $K$-distributions, any oblateness, and general dipole-moment configurations.

For the smallest grains ($a \lesssim 5\,\mathrm{\text{\AA}}$), which should dominate the AME, spontaneous rotational emission is a primary channel for angular-momentum loss and must be quantified accurately. A quantum-mechanical derivation of the spontaneous-emission rate for axisymmetric grains already exists in the \textsc{spdust} literature, but the rate is evaluated under the same simplifying assumption of homogeneous $K$-populations. This approximation is reasonable only when other torques acting on the grain are negligible on the timescale over which the $K$-sublevels are scrambled. The relative importance of spontaneous emission torques over the $K$-scrambling timescale is $\sim \frac{1}{J}\,\frac{A_{JK}}{\dot{N}_{\rm abs}},$ where $A_{JK}$ is a characteristic spontaneous-emission rate of state $\ket{JK}$, and $\dot{N}_{\rm abs}$ is the UV photon absorption rate. For grains smaller than $\sim 4\,\mathrm{\text{\AA}}$, one typically finds $A_{JK}/\dot{N}_{\rm abs}\sim 1$, implying that, in the absence of other torques, the grain can lose a large fraction of its angular momentum to spontaneous emission before the $K$-sublevels are fully redistributed. This matters because $A_{JK}$ depends strongly on $K$; where for the extremes $A_{J0}/A_{JJ} \simeq (1+\delta)^{-1}(\frac{\mu_{\parallel}^2}{\mu_{\perp}^2} + 1/4)$, implying significant variation. In this regime, the assumption of instantaneous (or effectively complete) $K$-scrambling can therefore become inaccurate---not only in the emissivity calculation, but also in determining the angular-momentum distribution that enters the \textsc{spdust} Fokker--Planck treatment.



\section{Conclusions}
\label{sec:conclusions}
We have developed a quantum-mechanical framework for the polarization of radiation from very small, axisymmetric dust grains, with the specific aim of assessing when their purely rotational emission (AME/spinning dust) can become measurably polarized. To achieve this, we have leveraged concepts from the established theory of atomic and molecular alignment in directional radiation fields. We quantify how anisotropic illumination by an interstellar radiation field produces alignment in individual rotating VSGs, and subsequently polarizes its emission. The polarization fraction of spinning dust emission is a function of (i) the radiation-field anisotropy $w$ (including its sign, set by illumination geometry relative to $\boldsymbol{B}$), (ii) the dipole-anisotropy parameter $\hat{\mu}_{\rm anis}$ (set by the transition dipole configuration), and (iii) the competition between alignment pumping by absorptions and depolarization by rotational spontaneous emission. Under optimal conditions, when VSGs are illuminated by a unidirectional radiation field in the direction of the magnetic field, we find polarization fractions near the maximum of the total emission of order $\sim 3\%$ for the purely parallel-dipole case and $\sim 1$--$1.5\%$ for perpendicular and shape-derived dipole configurations when averaging over a plausible VSG size distribution. These estimates will likely be a factor $10$ lower when considering relevant radiation anisotropies in diffuse molecular or atomic clouds. The polarization direction is with respect to the magnetic field direction, but can be parallel or perpendicular to its plane of the sky projection, depending on the dipole and radiation anisotropy directions. 

In addition to emission, the framework delivers self-consistent estimates of the polarized absorption coefficients for IR/optical/UV transitions from VSGs. We showed that the absorption polarization fraction scales approximately linearly with optical depth, $p_Q \simeq (\kappa_Q/\kappa_I)\,\tau_I$, with $\kappa_Q/\kappa_I$ set by the same anisotropic-illumination alignment physics that controls the microwave emissivity. This naturally identifies environments with large $|w|$ (e.g.\ strongly illuminated PDRs and reflection nebulae) as promising targets for joint constraints from AME polarization and PAH-band or UV/optical absorption polarimetry. In contrast to their absorption signatures, we predict negligible polarization for IR emission from VSGs aligned by anisotropic radiation.


\begin{acknowledgement}
BL acknowledges funding from the European Union’s Horizon Europe research and innovation programme under the Marie Skłodowska-Curie grant agreement No. 101126636.
\end{acknowledgement}
\bibliographystyle{aa}
\bibliography{lib.bib}

\begin{appendix}
\onecolumn
\section{Hamiltonian and eigenstates of an isolated very small grain} 
\subsection{Hamiltonian of an isolated very small grain}
We consider the Hamiltonian of the isolated very small grain (VSG). Fundamentally, the general Hamiltonian includes the kinetic energy of electrons and nuclei, as well as the Coulomb interactions between all particles \citep{szabo:12}. 
However, evaluating these interactions for a many-body VSG is prohibitively complex. To make modeling feasible, we apply a series of approximations.

First, we invoke the Born-Oppenheimer approximation, which assumes that electronic motions are much faster than nuclear motions \citep{papousek:82}. This implies that the nuclei move in a force field generated by an electronic structure that adapts instantaneously to the changing nuclear configuration. The nuclear motion comprises $3N$ degrees of freedom, where $N$ is the number of nuclei. These are categorized into translation (3 modes), rotation (3 modes), and vibration ($3N-6$ modes). We omit analysis of the translational modes, as these are trivial and can be treated separately. However, the same independence does not apply to rotational and vibrational motions; strictly speaking, they interact through mechanisms such as Coriolis coupling \citep{papousek:82}. 

The picture becomes even more complex when we consider that the rotating VSG is not a rigid body, but contains constituents with their own intrinsic angular momenta \citep{purcell:79}. Specifically, unpaired electrons possess orbital and spin angular momentum. Therefore, the conservation of angular momentum must encompass the coupled system of rotational, vibrational, and electronic momenta \citep{zare:88}. These electronic momenta couple to the particle's rotation through Coriolis terms and spin-rotation interactions (which arise from relativistic magnetic effects) \citep{brown:03}. Additionally, mutual interactions occur between the electrons themselves. To be comprehensive, one must also account for nuclear spin; nuclei couple to the bulk rotation, to unpaired electrons, and to other nuclei \citep{lazarian:99}. Collectively, the energy splittings caused by electron spin and orbital moments are known as fine structure, whereas the additional splittings due to nuclear spin are known as hyperfine structure \citep{brown:03}.

Based on the interactions described above, the effective Hamiltonian of the VSG is given by
\begin{align}
\label{eq:tot_ham}
\hat{H} = \hat{H}_{\mathrm{rot}} + \hat{H}_{\mathrm{vib}} + \hat{H}_{\mathrm{fs}} + \hat{H}_{\mathrm{hfs}} + \hat{C},
\end{align}
where, $\hat{H}_{\mathrm{rot}}$ is the rotational Hamiltonian, $\hat{H}_{\mathrm{vib}}$ is the vibrational Hamiltonian, $\hat{H}_{\mathrm{fs}}$ the fine-structure Hamiltonian (excluding rotation terms), $\hat{H}_{\mathrm{hfs}}$ the hyperfine-structure Hamiltonian (excluding rotation terms) and $\hat{C}$ the remaining terms that couple vibrational, fine-structure and hyperfine-structure terms to each other, and most importantly, to the rotation.

\subsection{Eigenstates of the isolated very small grain}
We now proceed to determine the eigenstates of the VSG. Given the coupling between all interactions, it is standard in quantum mechanical analysis to define the system in terms of the total angular momentum $\hat{\boldsymbol{F}}$, which is the vector sum of all constituent angular momenta \citep{landau:13},
\begin{align}
\label{eq:total_ang}
\hat{\boldsymbol{F}} = \hat{\boldsymbol{J}} + \sum_v \hat{\boldsymbol{l}}_v + \sum_i \left(\hat{\boldsymbol{\ell}}_i + \hat{\boldsymbol{s}}_i\right) + \sum_{\alpha} \hat{\boldsymbol{i}}_{\alpha},
\end{align}
where $\hat{\boldsymbol{J}}$ is the rotational angular momentum, $\hat{\boldsymbol{l}}_v$ is the angular momentum operator corresponding to vibrational mode $v$, $\hat{\boldsymbol{\ell}}_i$ and $\hat{\boldsymbol{s}}_i$ are the orbital and spin angular momentum operators of (unpaired) electron $i$, and $\hat{\boldsymbol{i}}_{\alpha}$ is the nuclear spin angular momentum operator of nucleus $\alpha$. It is evident that for a VSG, accounting for the coupling of such a large number of angular momentum operators is not a feasible task.

It is at this point that we start to let go of our general discussion and start to specialize to VSGs in the interstellar medium (ISM). It is then interesting to quantify the expected quanta of angular momenta, and with it perform an order-of-magnitude analysis on the terms in the Hamiltonian of Eq.~(\ref{eq:tot_ham}). We begin by considering rotational interactions. To an order of magnitude, the rotational energy of a VSG of size, $a$, and density $\rho$, is of the order, $E_{\mathrm{rot}} \sim \frac{15}{16 \pi} \frac{J^2}{\rho a^5},$
which at thermal energy, corresponds to 
\begin{subequations}
\label{eq:ang_quanta}
\begin{align}
\label{eq:J_typ}
J/\hbar \sim 2057 \ \left( \frac{T}{100 \ \mathrm{K}}\right)^{1/2} \left( \frac{\rho}{3 \ \mathrm{g/cm^3}}\right)^{1/2} \left( \frac{a}{10 \ \mathrm{\text{\AA}}}\right)^{5/2}, 
\end{align}
quanta of rotational angular momenta. Comparing the amount of rotational angular momentum quanta to the quanta due to electron and nuclear spins, we use that the amount of spin quanta (and orbital quanta), are proportional to the number of atoms in the VSG \citep{draine:16},
$$
N \sim 400 \left(\frac{a}{10\ \mathrm{\text{\AA}}}\right)^3.
$$
Only a fraction of these atoms are associated with electron- or nuclear spins,
\begin{align}
S_{\mathrm{max}}/\hbar \sim 40 \ \left( \frac{f_p}{0.1} \right) \left( \frac{a}{10 \ \mathrm{\text{\AA}}}\right)^{3}, \\
I_{\mathrm{max}}/\hbar \sim 20 \ \left( \frac{f_n}{0.05} \right) \left( \frac{a}{10 \ \mathrm{\text{\AA}}}\right)^{3}, 
\end{align}
\end{subequations}
where $f_p$ is the fraction of atoms that are paramagnetic, and $f_n$ is the fraction of nuclei that posses a non-zero nuclear spin. The values quoted above represent the maximum limit, corresponding to a fully polarized sample. However, since nanosized dust particles are typically not ferromagnetic, we do not expect full alignment; instead, the probable total spin angular momentum scales with the square root of these maxima. Finally, the VSG will also possess vibrational angular momentum quanta; while difficult to quantify precisely, we expect their number to be small compared to $J/\hbar$, as the excitation energy of vibrational modes substantially exceeds that of rotational modes.

Having evaluated the expected magnitudes of the different quanta of angular momentum, it is interesting to compare the scalings of the rotational Hamiltonian with the elements that couple rotation to other angular momenta. We have already established that rotational energy scales with the square of the rotational angular momentum. This can be contrasted to the interactions \citep{townes:55},
\begin{subequations}
\begin{align}
\hat{C}_{\mathrm{rv}} &\propto \hat{\boldsymbol{\ell}}_{v}\cdot \hat{\boldsymbol{J}}, \\
\hat{C}_{\mathrm{rs}} &\propto \hat{\boldsymbol{s}}_i\cdot \hat{\boldsymbol{J}}, \\ 
\hat{C}_{\mathrm{ri}} &\propto \hat{\boldsymbol{i}}_\alpha\cdot \hat{\boldsymbol{J}}, 
\end{align}
\end{subequations}
which couple the rotation to the vibration (Coriolis interaction), total spin (spin-rotation interaction), and total nuclear spin (nuclear spin-rotation interaction). All of these interactions scale linearly with the rotational angular momentum. Additionally, the coupling constants for these interactions, like the rotational constant, scale with the inverse moment of inertia \citep{brown:03}. Thus, for large quanta of rotational angular momentum, the interaction terms coupling to the rotational angular momentum lose their importance relative to the rotational Hamiltonian. When this is the case, it is a good approximation to consider the total wave function of the VSG as the product of a rotational wave function and a wave function describing all other motions and interactions: $\Psi = \psi_{\mathrm{\mathrm{rot}}} \psi_{\mathrm{rest}}$ \citep{landau:13}.

\section{The interaction of very small grains with magnetic fields}
In the main body of the paper, we operate under the assumption that the VSG aligns with and precesses rapidly around an ambient magnetic field. This precession arises from the interaction between the external magnetic field $\boldsymbol{B}$ and the VSG magnetic moment $\boldsymbol{m}$. Specifically, the field exerts a torque $\boldsymbol{\tau} = \boldsymbol{m} \times \boldsymbol{B}$ on the particle, driving the rotational angular momentum vector to gyrate around the field lines. The characteristic frequency of this precession is given by $\Omega_B \sim mB /J$ \citep{lazarian:07}.

The magnetic moment of interstellar dust and VSGs is commonly attributed to the Barnett effect, where particle rotation causes electron spins to align, producing a coherent magnetic moment \citep{purcell:79, andersson:15}. However, considering that relaxation is required for spins to align themselves along the rotation-induced magnetic field, and in the cold phase of the grain, these relaxation pathways are strongly suppressed. Also, in the instances that the grain is warm, during its thermal spikes, the grain rotation can exchange angular momentum with the vibrational modes, so that the magnetization may not be aligned with the rotation when the grain cools down to temperatures beyond which relaxation pathways are efficient.

We therefore describe how an additional magnetic moment complements the magnetic moment due to the Barnett effect. In order to do this, we draw inspiration from studies of molecular magnetism, particularly of closed-shell species (molecules with no unpaired electrons), which are known to acquire magnetic moments on the order of a nuclear magneton purely due to rotation \citep{flygare:71, flygare:78, steiner:01, wilson:05, lankhaar:18}. Physically, this occurs because the spatial distribution of the positive nuclear charge differs from that of the negative electronic cloud. Consequently, the electric currents generated by the rotating nuclei and electrons do not perfectly cancel, resulting in a net magnetic moment. We define this rotational magnetic moment for a collection of nuclei and electrons as \citep{jackson:98, lankhaar:18},
\begin{align}
\boldsymbol{m} = (\boldsymbol{w}^{\mathrm{nuc}} + \boldsymbol{w^{\mathrm{el}}}) \boldsymbol{I}^{-1} \boldsymbol{J},
\end{align}
where the coupling constants
\begin{subequations}
\begin{align}
\boldsymbol{w}^{\mathrm{nuc}}&= \sum_{\alpha} \frac{Q_\alpha}{4} \left[ \boldsymbol{r}_\alpha \cdot \boldsymbol{r}_\alpha  - \boldsymbol{r}_\alpha \boldsymbol{r}_\alpha^T \right], \\
\boldsymbol{w}^{\mathrm{el}}&= \sum_{i} \int d\boldsymbol{r}_i\ \frac{\rho_i(\boldsymbol{r}_i)}{4} \left[ \boldsymbol{r}_i \cdot \boldsymbol{r}_i \boldsymbol{1} - \boldsymbol{r}_i \boldsymbol{r}_i^T \right] ,
\end{align}
\end{subequations}
are functions of the nuclear (electron) positions, $\boldsymbol{r}_\alpha$ ($\boldsymbol{r}_i$), the nuclear charges $Q_\alpha$ and the electron charge density $\rho_i(\boldsymbol{r}_i)$, that is normalized to 
$$\int d\boldsymbol{r}_i\ \rho_i(\boldsymbol{r}_i) = -e.$$
To get an order-of-magnitude estimate for the rotational magnetic moment, we model the VSG as a sphere of radius $a$ containing $N$ randomly distributed nuclei. We assume each nucleus carries a charge $+e$ and is paired with an electron that is, on average, co-spatial with the nucleus but exhibits a positional spread corresponding to an effective atomic radius, $r_{\mathrm{atom}}$.  Under these assumptions, the coupling constant evaluates to
\begin{align}
\boldsymbol{w}^{\mathrm{nuc}} + \boldsymbol{w}^{\mathrm{el}} = - \frac{N e r_{\mathrm{atom}}^2}{6} \boldsymbol{1}.
\end{align}
Using the moment of inertia for a sphere, $I \sim \frac{2}{5} N m_{\mathrm{atom}} a^2$, we can derive the precession rate:
\begin{subequations}
\begin{align}
\Omega_B = \frac{|\boldsymbol{m}| B}{J} \sim \frac{5}{12}\frac{eB}{m_{\mathrm{atom}}} \left(\frac{r_{\mathrm{atom}}}{a}\right)^2.
\end{align}
This expression demonstrates a clear scaling of the rotational magnetic moment with the nuclear magneton. Taking a mean atomic weight of $21.6$ amu, a mean atomic size of $1\ \text{\AA}$, and a magnetic field of $10$ $\mathrm{\mu G}$, we calculate the precession rate as
\begin{align}
\Omega_B \sim 1.8 \times 10^{-5} \ \mathrm{s}^{-1}\  \left(\frac{a}{10\ \mathrm{\text{\AA}} } \right)^{-2} \left(\frac{B}{10\ \mathrm{\mu G} } \right)   \left(\frac{r_{\mathrm{atom}}}{1\ \mathrm{\text{\AA}} } \right)^{2}.
\end{align}
\end{subequations}
One might object that for heavier nuclei, many electrons exist as tightly bound core electrons confined to radii $\lesssim 0.5\ \mathrm{\text{\AA}}$ to the nucleus. However, the valence electrons in the outer shells, which form chemical bonds, often extend beyond $1\ \mathrm{\text{\AA}}$. This is particularly relevant for the aromatic compounds constituting graphitic grains, where electron delocalization can significantly increase the effective electron spread \citep{steiner:01}. Since the term $\boldsymbol{w}^{\mathrm{el}}$ depends on the square of the position, these outermost electrons dominate the magnetic moment. Based on this reasoning, and by comparison with the known rotational magnetic moments of large organic molecules, we are confident that our estimate is conservative.

We can compare the rotational Larmor precession rate derived above to the rate expected from the Barnett effect \citep{hoang:16},
\begin{align}
\Omega^{\mathrm{Barnett}} \sim 3 \times 10^{-3} \ \mathrm{s}^{-1} \ \left(\frac{a}{10 \ \mathrm{\text{\AA}}}\right)^{-2} \left(\frac{B}{10\ \mathrm{\mu G}} \right) \left(\frac{\chi}{10^{-4}} \right) .
\end{align}
While the standard Barnett estimate appears larger for typical parameters, in case paramagnetic species are present in the grain, it requires relaxation pathways to align spins with the rotation. In contrast, rotational Larmor precession operates universally for both silicate and carbonaceous grains and, crucially, does not require any time-dependent relaxation of spins to establish the moment. 

For carbonaceous grains specifically, the rotational magnetic moment is likely dominant. First, electron delocalization significantly increases the effective electron spread ($r_{\mathrm{atom}}$), enhancing the magnetic moment due to rotation \citep{steiner:01}. Second, these grains typically lack unpaired electrons; so the Barnett effect applies to nuclear spins that are associated with magnetic moments, $m_e/m_p$ weaker compared to electron spins \citep{lazarian:99}.

\section{Radiative interactions in the dipole approximation}
Leveraging the separation between rotational and internal wavefunctions established in the main body of the paper, the coupling between a unidirectional oscillating electric field (at resonance) and the dipole moment is given by \citep{landi:84},
\begin{align}
\label{eq:dipole}
&\braket{(\gamma K) J M \vert \boldsymbol{\mu}\cdot\boldsymbol{E} \vert (\gamma' K') J' M'}
=
\sum_{q=-1}^{+1} (-1)^q\,
\braket{(\gamma) J M \vert \mu_q \vert (\gamma') J' M'}  E_{-q} \nonumber
\\
&=
(-1)^{J-M}\,
\big\langle (\gamma K) J \big\| \mu \big\| (\gamma K') J' \big\rangle
\sum_{q=-1}^{+1} (-1)^q
\begin{pmatrix}
J & 1 & J' \\
-M & q & M'
\end{pmatrix} 
\sum_{\lambda} D_{\lambda q}^{(1)}(\Omega_{\boldsymbol{k}}) E_{\lambda}.
\end{align}
In the first line, we expanded the scalar product into spherical tensor components, $\boldsymbol{\mu}\cdot\boldsymbol{E} = \sum_{q=-1}^{+1} (-1)^q \mu_q E_{-q}$. In the second line, we applied the Wigner-Eckart theorem to factorize the matrix elements of the dipole operator $\mu_q$. This isolates the angular dependence (encoded in the Wigner 3j-symbol, in round brackets) from the structural dynamics, which are contained in the reduced dipole matrix element \citep{landi:84}
\[
\langle (\gamma K) J \| \mu \| (\gamma' K') J' \rangle.
\]
Finally, we related the electric field components in the space-fixed frame (index $q$) to the radiation-field frame (index $\lambda$) using Wigner rotation matrices, $D_{\lambda q}^{(1)}(\Omega_{\boldsymbol{k}})$. Here, $\lambda$ denotes the polarization states of the photon, and the angles $\Omega_{\boldsymbol{k}}$ describe the rotation between the space-fixed frame and the radiation propagation direction.

\subsection{Absorption rates for alignment elements}
Equation~(\ref{eq:dipole}) shows that the dipole–radiation interaction is inherently directional, with a strong dependence on the magnetic quantum number $M$ and on the specific transition. We now consider the interaction between the ISRF and a very small grain (VSG). For an isotropic ISRF, the $M$-dependence averages out, producing no net alignment. In contrast, the presence of an anisotropic radiation component renders the absorption rate sensitive to the grain alignment state.

To track imbalances among the $M$-sublevels, we employ the irreducible tensor operator formalism introduced in the main text. The treatment of anisotropic radiation fields and their impact on alignment populations is developed in \citet{landi:84} and in the monograph by \citet{landi:06}. Following \citet{landi:84}, the rate of change of the state multipoles due to absorption transitions to upper levels $\ket{\gamma' J' K'}$ is given by
\begin{align} 
\label{eq:landi_dep}
&\left[\dot{\rho}_{0}^k([\gamma K] J)\right]_{\mathrm{abs}}^{\mathrm{dep}} = -\sum_{\gamma'} \sum_{J'K'} B_{\gamma JK \to \gamma'J'K'} J_0^0(\nu_{\gamma JK \to \gamma'J'K'}) \sum_{k'} \rho_0^{k'}([\gamma K]J)\nonumber \\ 
&\times \left[ \delta_{kk'} + (-1)^{1 - J + J'} (2J+1)\sqrt{15 (2k+1)(2k'+1)} \left\{\begin{matrix} k & k' & 2 \\ J & J & J \end{matrix} \right\} \left\{\begin{matrix} 1 & 1 & 2 \\ J & J & J' \end{matrix} \right\} \left(\begin{matrix} k & k' & 2 \\ 0 & 0 & 0 \end{matrix} \right)\frac{J_0^2(\nu_{\gamma JK \to \gamma' J'K'})}{J_0^0(\nu_{\gamma JK \to \gamma'J'K'})}  \right] ,
\end{align}
where we consider the second irreducible element of the radiation field tensor, in the approximation of weakly polarized exciting radiation \citep{landi:06}
\begin{subequations}
\begin{align}
J_0^2 (\nu) = \frac{1}{4\pi}\int d\Omega_{\boldsymbol{k}} \ \frac{P_2(\boldsymbol{k}\cdot \boldsymbol{b})}{\sqrt{2}}I_{\nu} (\Omega_{\boldsymbol{k}}).
\end{align}
The second irreducible element of the radiation field tensor corresponds to the specific intensity integrated over all solid angles, weighted by the second Legendre polynomial of the angle between the radiation propagation direction $\boldsymbol{k}$ and the magnetic field direction $\boldsymbol{b}$. To quantify the degree of anisotropy, we define the radiation anisotropy factor:
\begin{align}
w(\nu) = \sqrt{2}\ J_0^2 (\nu)/ J_0^0 (\nu),
\end{align}
\end{subequations}
where we note that this factor can be positive or negative. In the subsequent analysis, we assume this factor is frequency-independent, such that $w(\nu)=w$.

We adopt the high-$J$ limit, to use the asymptotic limits of the factor, that occurs in Eq.~(\ref{eq:landi_dep}),
\begin{equation}
  F(k,k';J,J')
  \equiv
  (2J+1)\sqrt{15(2k+1)(2k'+1)}\,(-1)^{1+J-J'}
  \begin{Bmatrix}
    1 & 1 & 2\\
    J & J & J'
  \end{Bmatrix}
  \begin{Bmatrix}
    k & k' & 2\\
    J & J & J
  \end{Bmatrix}
  \begin{pmatrix}
    k & k' & 2\\
    0 & 0 & 0
  \end{pmatrix}.
  \label{eq:F_factor_def}
\end{equation}
In the large-$J$ limit, the relevant $F$ factors reduce to branch-dependent constants.
For $J'=J\pm 1$ (the $P/R$ branches) and $J'=J$ (the $Q$ branch), we take
\begin{align}
  F(0,2;J,J\pm 1) &= \sqrt{\frac{1}{10}},
  &F(0,2;J,J) &= -\sqrt{\frac{2}{5}},
  \nonumber\\
  F(2,2;J,J\pm 1) &= \frac{\sqrt{2}}{7},
  &F(2,2;J,J) &= -\frac{2\sqrt{2}}{7},
  \label{eq:F_largeJ_values}
\end{align}
and $F(0,0;J,J')=0$. Using these limits, we compute the depletion rate from Eqs.~(\ref{eq:einstein_symm}) and (\ref{eq:landi_dep}), to obtain
\begin{align}
\label{eq:rho2_abs_dep_largeJ_compact}
\left[\dot{\rho}_{0}^2([\gamma K] J)\right]_{\mathrm{abs}}^{\mathrm{dep}} &= -\dot{N}_{\mathrm{abs}}\left[ -\sqrt{\frac{1}{5}} w \hat{\mu}_{\mathrm{anis}}  P_2\left(\frac{K}{J}\right) \rho_0^0([\gamma K]J)+ \left[1 - \frac{2}{7} w \hat{\mu}_{\mathrm{anis}}  P_2\left(\frac{K}{J}\right) \right] \rho_0^2([\gamma K]J)\right],
\end{align}
where we have defined the dipole anisotropy parameter as:
\begin{align}
\hat{\mu}_{\mathrm{anis}}= \frac{|\mu_{\parallel}|^2 - \frac{1}{2}|\mu_{\perp}|^2}{|\mu_{\parallel}|^2 + |\mu_{\perp}|^2}.
\end{align}
This term dictates how strongly the absorption probability depends on the particle's orientation.

Conversely, an absorption event promotes the VSG to an upper energy state. The corresponding pumping rate into the upper alignment level is \citep{landi:84}
\begin{align}
\left[\dot{\rho}_{0}^2([\gamma' K'] J')\right]_{\mathrm{abs}}^{\mathrm{pump}} &= \dot{N}_{\mathrm{abs}} \left[ -\sqrt{\frac{1}{5}} w  \hat{\mu}_{\mathrm{anis}}  P_2\left(\frac{K}{J}\right) \rho_0^0([\gamma K]J) + \left[1 - \frac{2}{7} w \hat{\mu}_{\mathrm{anis}}  P_2\left(\frac{K}{J}\right) \right] \rho_0^2([\gamma K]J) \right].
\end{align}
In deriving this pumping term, we assumed that the alignment populations vary smoothly between adjacent rotational states (i.e., $\rho_0^2$ is similar between $\ket{JK}$ and nearby states).

\subsection{Depolarization rate due to spontaneous emission}
We proceed to consider the impact of spontaneous emission processes on the alignment states. As we did for absorption processes, we divide the rate of change of a level, $\ket{(\gamma) JK}$, alignment element, into processes that deplete the level, due to spontaneous emission events from higher states \citep{landi:84},
\begin{align}
\left[ \dot{\rho}_0^{2}([\gamma K] J) \right]_{\mathrm{spont}}^{\mathrm{depl}} = \sum_{\gamma'} \sum_{J'K'} A_{\gamma J K \to \gamma' J' K' }\  \rho_0^{2}([\gamma K] J),
\end{align}
where the summation is over all states, $\ket{\gamma' J' K'}$ that are lower in energy than the $\ket{(\gamma) JK}$ level. Conversely, the pumping of the level $\ket{(\gamma) JK}$, due to spontaneous emission processes, is
\begin{align}
\left[ \dot{\rho}_0^{2}([\gamma K] J) \right]_{\mathrm{spont}}^{\mathrm{pump}} = \sum_{\gamma'} \sum_{J'K'} (-1)^{1 + J + J'} (2J+1) \begin{Bmatrix} J & J & k \\ J' & J' & 1\end{Bmatrix}A_{\gamma' J' K' \to \gamma J K}\  \rho_0^{k}([\gamma' K'] J') ,
\end{align}
where the summation is over all states, $\ket{\gamma' J' K'}$ that are higher in energy than the $\ket{(\gamma) JK}$ level.

These expressions allow us to  estimate the depolarization of spinning dust grains, through their purely rotational emission. In case of purely rotational emission, only Q-branch $J \to J$ and P-branch $J\to J-1$ transitions are allowed, of which the P-branch transition are dominant (since the Einstein coefficient scales with the frequency cubed). We can evaluate the Wigner 6j-symbol in the limit of high-$J$, 
\[
(2J+1)\begin{Bmatrix}J & J & 2 \\ J+1 & J+1 & 1 \end{Bmatrix} \approx 1 - \frac{1}{2J},
\]
so that the rate of change of an alignment level due to pure rotational emission is
\begin{align}
\left[ \dot{\rho}_0^{2}([\gamma K] J) \right]_{\mathrm{spont}}^{\mathrm{depol}} &\simeq - \rho_{0}^{2}([\gamma K]J)\sum_{K'} A_{J K \to J-1 K'}  \nonumber \\ &+ \sum_{K'} A_{J+1 K' \to J K} \left(1 - \frac{1}{2J} \right) \rho_0^{2} ([\gamma K']J+1).
\end{align}
We can turn this rate into an effective depolarization rate by making the approximations
\[
\frac{A_{J+1 K' \to J K}}{A_{JK' \to JK}} \approx \left(\frac{2B(J+1)}{2BJ}\right)^3 \simeq 1 + \frac{3}{J},
\]
for the ratio of the Einstein coefficients, and
\[
\frac{\rho_0^{2}([\gamma K']J+1)}{\rho_0^{2} ([\gamma K]J)} \approx \sqrt{\frac{2J+3}{2J+1}} e^{-(2 h B/kT_{\mathrm{rot}})(J+1)} \simeq 1 + \frac{1}{2J} - \frac{2hB}{kT_{\mathrm{rot}}}J,
\]
for the alignment populations of the upper and lower levels. These approximations amount to the assumptions that (i) all transition center around the P-branch $\Delta K =0$ transitions, (ii) the level populations are thermally distributed at temperature $T_{\mathrm{rot}}$, and (iii) the relative alignment of adjacent rotational levels is the same. Adopting these assumptions allows us to estimate an effective depolarization rate of state $\ket{\gamma JK}$, due to rotational emission,
\begin{align}
&\left[ \dot{\rho}_0^{2}([\gamma K] J) \right]_{\mathrm{spont}}^{\mathrm{depol}} \approx \sum_{K'}- A_{JK \to J-1 K} \rho_0^{2}([\gamma K]J) \nonumber \\ &\times \left[1 - \left(\left[1-\frac{2}{J}\right] \left[\frac{J+1}{J}\right]^3 \frac{\rho_0^2 ([\gamma K] J+1)}{\rho_0^2 ([\gamma K] J)} \right) \right]  =-A^2_0 \rho_0^{2},
\end{align}
where
\begin{align}
\label{eq:spont_depol}
A^2_0 = \left(\frac{2hB}{kT_{\mathrm{rot}}}J - \frac{5}{2J} \right)\sum_{K'} A_{JK \to J-1 K'}.
\end{align}
We note that for the very small dust grains, depolarization is small, as the $J$-states that are excited are typically high (see Eq.~\ref{eq:J_typ}). Note furthermore, that with $hBJ_{\mathrm{th}}^2 \sim kT$, so the term $\frac{2hB}{kT_{\mathrm{rot}}}J \propto J^{-1}$. 

\section{Rate equations for VSGs in anisotropic radiation fields}
We are now in a position to synthesize the individual processes derived above---anisotropic absorption, relaxation, and pumping---into a set of coupled rate equations that describe the time evolution of the alignment populations. We combine these terms to evaluate the total rate of change for the alignment populations of the VSG. We make a rough divide into alignment populations in the ground state, and an excited state, that is populated immediately after an absorption event. Let
$
  \rho^2_{0,g} \equiv \rho^{2}_{0}\!\left([\gamma K]J\right),
  $ and $
  \rho^0_{0,g} \equiv \rho^{0}_{0}\!\left([\gamma K]J\right).
  $
To model their time-evolution, we include three processes:
(i) absorption depletion of the ground-level alignment, (ii) decay from the excited to ground manifold at rate $\Gamma$, and (iii) intrinsic ground-state alignment relaxation at rate $A_0^2$.
The alignment rate equations are then taken to be
\begin{subequations}
\begin{align}
  \dot{\rho}^{2}_{0,g}
  &=
  \left[\dot{\rho}^{2}_{0,g}\right]_{\mathrm{abs}}^{\mathrm{dep}}
  + \Gamma\,\rho^{2}_{0,e}
  - A_0^2\,\rho^{2}_{0,g},
  \label{eq:rho2g_two_level_rate}\\[4pt]
  \dot{\rho}^{2}_{0,e}
  &=
  -\left\langle\left[\dot{\rho}^{2}_{0,g}(K)\right]_{\mathrm{abs}}^{\mathrm{dep}}\right\rangle_{K}
  - \Gamma\,\rho^{2}_{0,e},
  \label{eq:rho2e_two_level_rate}
\end{align}
\end{subequations}
where Eq.~(\ref{eq:rho2e_two_level_rate}) expresses that the absorption pumping of the excited-state per-$K$ alignment is the negative of the total ground depletion, averaged over the excited $K$ levels. The averaging is due to the fast scrambling of the excited state levels since relaxation is fast.

We seek the steady-state solution for the fractional ground alignment,
\begin{equation}
  \sigma^{2}_{0,g} \equiv \sigma^{2}_{0}\!\left([\gamma K]J\right) = \frac{\rho^{2}_{0,g}}{\rho^{0}_{0,g}}.
  \label{eq:sigma2g_def}
\end{equation}
Setting $\dot\rho^2_{0,e}=0$ in Eq.~(\ref{eq:rho2e_two_level_rate}) gives
\begin{equation}
  \Gamma\,\rho^{2}_{0,e}
  =
  -\left\langle\left[\dot{\rho}^{2}_{0,g}(K)\right]_{\mathrm{abs}}^{\mathrm{dep}}\right\rangle_{K}.
  \label{eq:rho2e_quasisteady_from_dep}
\end{equation}
Inserting Eq.~(\ref{eq:rho2e_quasisteady_from_dep}) into Eq.~(\ref{eq:rho2g_two_level_rate}) and setting $\dot\rho^2_{0,g}=0$ yields
\begin{equation}
  A_0^2\,\rho^{2}_{0,g}
  =
  \left[\dot{\rho}^{2}_{0,g}\right]_{\mathrm{abs}}^{\mathrm{dep}}
  -\left\langle\left[\dot{\rho}^{2}_{0,g}(K)\right]_{\mathrm{abs}}^{\mathrm{dep}}\right\rangle_{K}.
  \label{eq:rho2g_steady_balance}
\end{equation}
%
Writing Eq.~(\ref{eq:rho2g_steady_balance}) using Eq.~(\ref{eq:rho2_abs_dep_largeJ_compact}) gives, for each $K$,
\begin{equation}
  \left[A_0^2+\dot{N}_{\mathrm{abs}}\!\left(1+w\,c_{22}(K)\right)\right]\rho^{2}_{0,g}(K)
  =
  -\dot{N}_{\mathrm{abs}}\,w\,c_{02}(K)\,\rho^{0}_{0,g}(K)
  - \left\langle\left[\dot{\rho}^{2}_{0,g}(K)\right]_{\mathrm{abs}}^{\mathrm{dep}}\right\rangle_{K},
  \label{eq:rho2g_steady_K_explicit}
\end{equation}
where $c_{02}(K) = \sqrt{1/5}\hat{\mu}_{\rm anis} P_2 (K/J)$ and $c_{22}(K) = 2/7 \hat{\mu}_{\rm anis} P_2(K/J)$.

Taking the $K$-average of Eq.~(\ref{eq:rho2g_steady_balance}) implies $\left\langle \rho^{2}_{0,g}(K)\right\rangle_K=0$ in steady state (for $A_0^2$ independent of $K$).
We can therefore solve Eq.~(\ref{eq:rho2g_steady_K_explicit}) by treating the $K$-averaged depletion as an (unknown) constant and imposing $\langle\rho^2_{0,g}\rangle_K=0$:
\begin{subequations} 
  \label{eq:avg_dep_closed_form}
\begin{align}
  \rho^{2}_{0,g}(K)
  &=
  \frac{
    -\dot{N}_{\mathrm{abs}}\,w\,c_{02}(K)\,\rho^{0}_{0,g}(K)
    - \left\langle\left[\dot{\rho}^{2}_{0,g}(K)\right]_{\mathrm{abs}}^{\mathrm{dep}}\right\rangle_{K}
  }{A_0^2+\dot{N}_{\mathrm{abs}}\left(1+w\,c_{22}(K)\right)},
 \\
  \left\langle\left[\dot{\rho}^{2}_{0,g}(K)\right]_{\mathrm{abs}}^{\mathrm{dep}}\right\rangle_{K}
  &=
  -\dot{N}_{\mathrm{abs}}\,w\,
  \frac{
    \left\langle \dfrac{c_{02}(K)\,\rho^{0}_{0,g}(K)}{A_0^2+\dot{N}_{\mathrm{abs}}\left(1+w\,c_{22}(K)\right)}\right\rangle_{K}
  }{
    \left\langle \dfrac{1}{A_0^2+\dot{N}_{\mathrm{abs}}\left(1+w\,c_{22}(K)\right)}\right\rangle_{K}
  }.
\end{align}
\end{subequations}
Equations~(\ref{eq:avg_dep_closed_form}) provide a closed-form solution for the ground-state alignment pattern across $K$ under fast K-scrambling in the excited state. The total population elements of the ground state, $\rho^{0}_{0,g}(K)$, are determined from Eq.~(\ref{eq:rot_distr}) as explained in the text.

\section{Emission and absorption terms in the continuum approximation}
We compute the emissivities $\eta_{I}$ and $\eta_{Q}$ of the Stokes $I$ and $Q$ components of the radiation field. Under the assumption of fast magnetic precession, we can put emissivities in the Stokes $U$ and $V$ components to zero. We choose the quantization axis along the alignment axis (i.e. the magnetic field direction), and let $(\vartheta,\chi)$ be the polar angles of the line of sight relative to this axis. For a given upper level $u\equiv([\gamma K]J)$, we defined the population tensor $\rho^{0}_{0}(u)$ and fractional alignment $\sigma^{2}_{0}(u)$. The irreducible tensor element of the population, relates to the total population probability of the $u$-level by
\begin{equation}
  P_u(a) = \sqrt{2J+1}\,\rho^{0}_{0}\!\left(u;a\right).
\end{equation}
The emissivity per grain into Stokes $I$ and $Q$ for a single line $u\to l$ can be written as
\begin{subequations}
\begin{align}
  \eta_{I}^{(u\to l)}(a;\vartheta)
  &=
  \frac{h\nu_{ul}}{4\pi}\,A_{u\to l}\,
  P_{u}(a)\,
  \phi\!\left(\nu-\nu_{ul}\right)\,
  \left[
    1 + w^{(2)}_{u l}\,\sigma^{2}_{0}(u)\,\mathcal{T}^{2}_{0}(I;\vartheta)
  \right],
  \label{eq:eta_I_line}\\[4pt]
  \eta_{Q}^{(u\to l)}(a;\vartheta)
  &=
  \frac{h\nu_{ul}}{4\pi}\,A_{u\to l}\,
  P_{u}(a)\,
  \phi\!\left(\nu-\nu_{ul}\right)\,
  \left[
    w^{(2)}_{u l}\,\sigma^{2}_{0}(u)\,\mathcal{T}^{2}_{0}(Q;\vartheta)
  \right],
  \label{eq:eta_Q_line}
\end{align}
\end{subequations}
where $w^{(2)}_{u l}$ is the (line-dependent) rank-2 polarizability factor (defined below), and the relevant geometrical tensors are
\begin{equation}
  \mathcal{T}^{2}_{0}(I;\vartheta)=\frac{1}{2\sqrt{2}}\left(3\cos^2\vartheta-1\right),
  \qquad
  \mathcal{T}^{2}_{0}(Q;\vartheta)=-\frac{3}{2\sqrt{2}}\sin^2\vartheta.
  \label{eq:T20_IQ}
\end{equation}
In the large-$J$ limit for the $P$ and $Q$ branches (i.e. $\Delta J=-1$ and $\Delta J=0$), one has $w^{(2)}_{J,J-1}\xrightarrow[J\to\infty]{}\sqrt{\frac{1}{10}}$ and 
  $w^{(2)}_{J,J}\xrightarrow[J\to\infty]{}-\sqrt{\frac{2}{5}}$. Summing over all lines gives the total and polarized emissivities per grain,
\begin{equation}
  \eta_{I}(a;\vartheta)=\sum_{u\to l}\eta_{I}^{(u\to l)}(a;\vartheta),
  \qquad
  \eta_{Q}(a;\vartheta)=\sum_{u\to l}\eta_{Q}^{(u\to l)}(a;\vartheta).
  \label{eq:eta_IQ_sum_lines}
\end{equation}
Finally, integrating over the size distribution yields the total (volume) emissivities
\begin{equation}
  \eta_{I}(\vartheta)=\int d a\,\frac{d n}{d a}\,\eta_{I}(a;\vartheta),
  \qquad
  \eta_{Q}(\vartheta)=\int d a\,\frac{d n}{d a}\,\eta_{Q}(a;\vartheta).
  \label{eq:jnu_IQ_integrated}
\end{equation}
Starting from the line emissivities Eqs.~(\ref{eq:eta_I_line}) and (\ref{eq:eta_Q_line}) and the sum Eq.~(\ref{eq:eta_IQ_sum_lines}), we derive branch-by-branch continuum approximations to the frequency-resolved emissivities in the high-$J$ limit.
We specialize to pure rotational (spinning-dipole) emission within a fixed vibronic manifold $\gamma$.
In the continuum approximation, we take the intrinsic profiles to be the Dirac delta function
\begin{equation}
  \phi\!\left(\nu-\nu_{ul}\right)\to \delta_{\mathrm{D}}\!\left(\nu-\nu_{ul}\right).
\end{equation}
Furthermore, for simplicity, formulating the unpolarized emissivity we neglect the rank-2 angular correction in Eq.~(\ref{eq:eta_I_line}),
\begin{subequations}
\begin{equation}
  \eta_{I}^{(u\to l)}(a)
  \simeq
  \frac{h\nu_{ul}}{4\pi}\,A_{u\to l}\,P_u(a)\,
  \delta_{\mathrm{D}}\!\left(\nu-\nu_{ul}\right),
  \label{eq:etaI_line_delta_noalign}
\end{equation}
whereas for Stokes $Q$ we retain Eq.~(\ref{eq:eta_Q_line}):
\begin{equation}
  \eta_{Q}^{(u\to l)}(a;\vartheta)
  =
  \frac{h\nu_{ul}}{4\pi}\,A_{u\to l}\,P_u(a)\,
  \delta_{\mathrm{D}}\!\left(\nu-\nu_{ul}\right)\,
  w^{(2)}_{u l}\,\sigma^{2}_{0}(u)\,\mathcal{T}^{2}_{0}(Q;\vartheta).
  \label{eq:etaQ_line_delta}
\end{equation}
\end{subequations}
For pure rotational emission the upper levels are $u\equiv([\gamma K]J)$.
The second part of the continuum approximation is that we treat $J$ as a continuous variable and replace the discrete $K$ sum by an integral over $x\equiv K/J$:
\begin{equation}
  \sum_{J=0}^{\infty}\ \sum_{K=-J}^{J} f(J,K)
  \xrightarrow[J\gg 1]{}
  \int_{0}^{\infty}d J\, J\int_{-1}^{1}d x\, f\!\left(J,Jx\right),
  \label{eq:JK_continuum_sum_rule}
\end{equation}
which is an approximation that increases its quality in proportion to the magnitude of $J$.

Integrals over the Dirac delta function, $\delta_{\mathrm{D}}(\nu-\nu(x))$, are made by using the substitution $\delta_{\mathrm{D}}(\nu-\nu(x))=\delta_{\mathrm{D}}(x-x_\star)/\left|d\nu/d x\right|_{x_\star}$, where $x_\star$ is the relevant solution. Since we keep ourselves to $u\equiv([\gamma K]J)$, we note from here on $P_u(a)=P (J,K;a)$ and alignments $\sigma^2_0(u)=\sigma^2_0(J,K;a)$. Now, we compute the emissivities per emission branch.
\subsubsection*{$P$-branch emissivities ($\Delta J=-1$)}
The dominant relaxing transitions have $\Delta J=-1$ (``P branch'') with the parallel and perpendicular channels corresponding to $\Delta K=0$ ($\propto \mu_{\parallel}^2$) and $\pm1$ ($\propto \mu_{\perp}^2$) transitions, as discussed in the main body. 

\paragraph*{(i) Parallel P branch ($\Delta K=0$).}
The parallel P-branch frequency is independent of $K$: $\nu^{(\parallel)}_{J,K}=2BJ$.
Using Eqs.~(\ref{eq:etaI_line_delta_noalign}- \ref{eq:JK_continuum_sum_rule}) gives
\begin{subequations}
\begin{equation}
  \eta_{I}^{(P,\parallel)}(a)
  \simeq
  \frac{8\pi^3}{3c^3}\,
  \left|\mu_{\parallel}\right|^2\,
  \nu^4\,
  \frac{J_\nu}{2B}
  \int_{-1}^{1}d x\,\Big[1-x^2\Big]\,
  P\!\left(J_\nu,J_\nu x;a\right),
  \label{eq:etaI_P_parallel_continuum_general}
\end{equation}
where $J_\nu = \frac{\nu}{2B}$ and the $\delta_{\mathrm{D}}(\nu-2BJ)$ factor has been integrated over $J$. Similarly, the continuum limit for the polarized intensity yields
\begin{equation}
  \eta_{Q}^{(P,\parallel)}(a;\vartheta)
  \simeq
  \frac{8\pi^3}{3c^3}\,
  \left|\mu_{\parallel}\right|^2\,
  \nu^4\,
  \mathcal{T}^{2}_{0}(Q;\vartheta)\,
  \sqrt{\frac{1}{10}}\,
  \frac{J_\nu}{2B}
  \int_{-1}^{1}d x\,\Big[1-x^2\Big]\,
  P\!\left(J_\nu,J_\nu x;a\right)\,
  \sigma^{2}_{0}\!\left(J_\nu,J_\nu x;a\right).
  \label{eq:etaQ_P_parallel_continuum_general}
\end{equation}
\end{subequations}
\paragraph*{(ii) Perpendicular P branch ($\Delta K=\pm 1$).}
In the large-$J$ limit, with $\delta=(B-A)/B$, the perpendicular P-branch frequencies are $\nu = 2BJ(1\pm \delta x)$. So for fixed $(\nu,J)$ it is convenient to define
\begin{equation}
  x_\nu(J)\equiv \frac{\nu/(2BJ)-1}{\delta}, 
  \label{eq:xnu_Pperp_def}
\end{equation}
Since we defined $K/J=x$, it obvious that $|x|<1$, and by extension $x_{\nu}$. Therefore, $\nu$ receives contributions only from
\begin{equation}
  \frac{\nu}{2B(1+\delta)}\le J \le \frac{\nu}{2B(1-\delta)}.
  \label{eq:J_bounds_Pperp_at_nu_generalPJK}
\end{equation}
Now taking the continuum limit of Eqs.~(\ref{eq:etaI_line_delta_noalign}), while integrating over $x$ with the delta functions for the two channels yields
\begin{subequations}
\begin{align}
  \eta_{I}^{(P,\perp)}(a)
  &\simeq
  \frac{\pi^3}{3 B c^3}\,
  \left|\mu_{\perp}\right|^2\,
  \frac{\nu^4}{\delta}
  \int_{\nu/[2B(1+\delta)]}^{\nu/[2B(1-\delta)]}\!\!d J\,
  \Big[1-x_\nu(J)\Big]^2\,
  \Big[
    P\!\left(J,Jx_\nu(J);a\right)
    +
    P\!\left(J,-Jx_\nu(J);a\right)
  \Big],
  \label{eq:etaI_P_perp_continuum_general}
\end{align}
where the first line keeps the channel-specific large-$J$ H\"onl--London weights from Eq.~(\ref{eq:einstein_symm}): $(1-x)^2$ for $\Delta K=+1$ evaluated at $x=x_\nu(J)$ and $(1+x)^2$ for $\Delta K=-1$ evaluated at $x=-x_\nu(J)$.
Both reduce to the common factor $[1-x_\nu(J)]^2$ because the two channels contributing at the same observed $\nu$ originate from opposite $K$ (opposite $x$).
The Jacobian factors $\left|d\nu/d x\right|^{-1}$ from integrating the delta functions are equal for the two channels and therefore introduce no additional relative $\pm$ sign.
The continuum limit for the polarized emissivity Eq.~(\ref{eq:etaQ_line_delta}) gives
\begin{align}
  \eta_{Q}^{(P,\perp)}(a;\vartheta)
  &\simeq
  \frac{\pi^3}{3 B c^3}\,
  \left|\mu_{\perp}\right|^2\,
  \frac{\nu^4}{\delta}\,
  \mathcal{T}^{2}_{0}(Q;\vartheta)\,
  \sqrt{\frac{1}{10}}
  \int_{\nu/[2B(1+\delta)]}^{\nu/[2B(1-\delta)]}\!\!d J\,
  \Big[1-x_\nu(J)\Big]^2
  \nonumber\\
  &\qquad\times
  \Big[
    P\!\left(J,Jx_\nu(J);a\right)\,\sigma^{2}_{0}\!\left(J,Jx_\nu(J);a\right)
    +
    P\!\left(J,-Jx_\nu(J);a\right)\,\sigma^{2}_{0}\!\left(J,-Jx_\nu(J);a\right)
  \Big].
  \label{eq:etaQ_P_perp_continuum_general}
\end{align}
\end{subequations}
\subsubsection*{$Q$-branch emissivities ($\Delta J=0$, $\Delta K=\pm 1$)}
The $Q$-branch transitions at fixed $J$ contribute at frequencies of order $(B-A)K$.
In the large-$J$ limit we use the approximation $\nu\simeq 2B\delta\,|K|$ and define
\begin{equation}
  K_\nu \equiv \frac{\nu}{2B\delta}.
  \label{eq:Knu_Qperp_def}
\end{equation}
Within the continuum approximation, a given $\nu$ then receives contributions from all $J\ge K_\nu$ with $K=\pm K_\nu$.

Using the large-$J$ perpendicular $Q$-branch factor Eq.~(\ref{eq:einstein_symm}) together with Eq.~(\ref{eq:etaI_line_delta_noalign}) in the continuum approximation, gives the unpolarized emissivity
\begin{subequations}
\begin{equation}
  \eta_{I}^{(Q,\perp)}(a)
  \simeq
  \frac{2\pi^3}{3 B c^3}\,
  \left|\mu_{\perp}\right|^2\,
  \frac{\nu^4}{\delta}
  \int_{K_\nu}^{\infty}d J\,
  \left[1-\left(\frac{K_\nu}{J}\right)^2\right]\,
  \Big[
    P(J,K_\nu;a)+P(J,-K_\nu;a)
  \Big],
  \label{eq:etaI_Q_perp_continuum_general}
\end{equation}
and the polarized emissivity is
\begin{align}
  \eta_{Q}^{(Q,\perp)}(a;\vartheta)
  &\simeq
  \frac{2\pi^3}{3 B c^3}\,
  \left|\mu_{\perp}\right|^2\,
  \frac{\nu^4}{\delta}\,
  \mathcal{T}^{2}_{0}(Q;\vartheta)\,
  \left(-\sqrt{\frac{2}{5}}\right)
  \int_{K_\nu}^{\infty}d J\,
  \nonumber \\ &\times \left[1-\left(\frac{K_\nu}{J}\right)^2\right]\,
  \Big[
    P(J,K_\nu;a)\,\sigma^{2}_{0}(J,K_\nu;a)
    +
    P(J,-K_\nu;a)\,\sigma^{2}_{0}(J,-K_\nu;a)
  \Big].
  \label{eq:etaQ_Q_perp_continuum_general}
\end{align}
\end{subequations}
\end{appendix}

\end{document}